\documentclass[12pt,draftclsnofoot,onecolumn,journal,comsoc]{IEEEtran}

\usepackage{cite}
\usepackage[T1]{fontenc}
\usepackage[pdftex]{graphicx}
\DeclareGraphicsExtensions{.pdf,.png}
\usepackage[cmex10]{amsmath}
\usepackage{amssymb}
\usepackage{amsthm}
\usepackage[cmintegrals]{newtxmath}
\usepackage{bm}
\usepackage{algorithmicx}
\usepackage{array}
\usepackage{mdwmath}
\usepackage{mdwtab}
\usepackage[caption=false,font=footnotesize]{subfig}
\usepackage{fixltx2e}
\usepackage{url}

\newtheorem{theorem}{Theorem}
\newtheorem*{theorem*}{Theorem}

\newtheorem{lemma}{Lemma}

\newtheorem{criterion}{Criterion}

\newcommand{\dif}{\mathop{}\!\mathrm{d}}

\DeclareMathOperator{\trace}{Tr}

\newcommand{\ket}[1]{\left| #1 \right>} 
\newcommand{\bra}[1]{\left< #1 \right|} 

\newcommand{\im}{j}

\newcommand{\BPSK}{\rm{b}}
\newcommand{\QPSK}{\rm{q}}
\newcommand{\ro}{\hat{\rho}_0} 
\newcommand{\rz}{{\hat{\rho}_{1,\BPSK}}} 
\newcommand{\ra}{\hat{\rho}_{00}} 
\newcommand{\rb}{\hat{\rho}_{01}} 
\newcommand{\rc}{\hat{\rho}_{10}} 
\newcommand{\rd}{\hat{\rho}_{11}} 
\newcommand{\nt}{\bar{n}_{\rm{B}}} 
\renewcommand{\d}{\mathrm{d}}
 
\newcommand{\ah}{\hat{a}} 
\newcommand{\ad}{\hat{a}^\dagger} 
\newcommand{\kh}{\hat{K}_{\BPSK}}

\begin{document}
\title{
	Fundamental limits of quantum-secure covert communication over bosonic channels 
}

\author{Michael~S.~Bullock,~\IEEEmembership{Student Member,~IEEE,}
        Christos~N.~Gagatsos,
        Saikat~Guha,~\IEEEmembership{Senior Member,~IEEE,}
        ~and Boulat~A.~Bash,~\IEEEmembership{Member,~IEEE}
\thanks{M.~S.~Bullock is with the Electrical and Computer Engineering Department, University of Arizona, Tucson, AZ. S.~Guha and B.~A.~Bash are with the Electrical and Computer Engineering Department, and the College of Optical Sciences, University of Arizona, Tucson, AZ. C.~N.~Gagatsos is with the College of Optical Sciences, University of Arizona, Tucson, AZ}
\thanks{CNG acknowledges the Office of Naval Research (ONR) MURI program on Optical Computing under grant no. N00014-14-1-0505. SG and BAB acknowledge the ONR program Communications and Networking with Quantum Operationally-Secure Technology for Maritime Deployment (CONQUEST), awarded under Raytheon BBN Technologies prime contract number N00014-16-C-2069, and subcontract to University of Arizona.}
\thanks{Some results from this manuscript were presented at the Central European Workshop on Quantum Optics (CEWQO) 2019.}}
\maketitle
\thispagestyle{empty} 
\vspace{-0.3in}
\begin{abstract}
We investigate the fundamental limit of quantum-secure covert communication over 
  the lossy thermal noise bosonic channel, the quantum-mechanical model
  underlying many practical channels.
We assume that the adversary has unlimited quantum information processing
  capabilities as well as access to all transmitted photons that
  do not reach the legitimate receiver.
Given existence of noise that is uncontrolled by the adversary, the
  \emph{square root law} (SRL) governs covert communication: up to $c\sqrt{n}$  
  covert bits can be transmitted reliably 
  in $n$ channel uses. 
Attempting to surpass this limit results in detection
  with unity probability as $n\rightarrow\infty$. 
Here we present the expression for $c$, characterizing the SRL for the bosonic 
  channel.
We also prove that discrete-valued coherent state quadrature phase shift keying
  (QPSK) constellation
  achieves the optimal $c$, which is the same as that achieved by 
  a circularly-symmetric complex-valued Gaussian prior on coherent state amplitude.
Finally, while binary phase shift keying (BPSK) achieves the Holevo capacity for 
  non-covert bosonic
  channels in the low received signal-to-noise ratio regime, we show that it is
  strictly sub-optimal for covert communication.
\end{abstract}
\IEEEpeerreviewmaketitle
\section{Introduction}
Covert, or low probability of detection/intercept (LPD/LPI), communication prevents 
  transmission's detection by an adversary.
This is a stricter security requirement than protection of transmission's content 
  from unauthorized access provided by the standard methods, e.g., encryption and
  quantum key distribution (QKD).
While covert communication has many practical applications, its fundamental limits
  were underexplored until \cite{bash12sqrtlawisit,bash13squarerootjsacnonote}
  proved that \emph{square root law} (SRL) governs covert communication over 
  additive white Gaussian noise (AWGN) channel: no more 
  than $c\sqrt{n}$ covert bits can be transmitted with arbitrarily small
  decoding error probability to the intended 
  receiver in $n$ uses of the channel, where $c$ is a constant and $n=TW$ is 
  the product of the transmission duration $T$ (in seconds) and the
  bandwidth $W$ (in Hz) of the source around its center frequency.
Attempting to transmit more results in either detection by the adversary with high 
  probability as $n\rightarrow\infty$, or unreliable transmission.
Even though the capacity of the covert channel is zero (since 
  $\lim_{n\rightarrow\infty}\frac{c\sqrt{n}}{n}=0$), as $n$ increases, 
  SRL allows transmission of a significant number of covert bits for large $n$.
Subsequent work extended
  \cite{bash12sqrtlawisit,bash13squarerootjsacnonote} by characterizing $c$
  \cite{bloch15covert,wang15covert}, showing the SRL for discrete memoryless
  channels (DMCs) \cite{che13sqrtlawbscisit,bloch15covert,wang15covert}, and
  determining it up to the second order \cite{tahmasbi19covertdmc2ndorder}.
A tutorial explanation of the SRL and its implications is offered in
  \cite{bash15covertcommmag}.
Consider an optical channel with additive thermal noise.
The use of laser light modulation at the transmitter and coherent detection 
  (homodyne or heterodyne) at the receiver induces an AWGN channel, with covert
  communication governed by the SRL in
  \cite{bash12sqrtlawisit,bash13squarerootjsacnonote}.
Fundamentally, however, electromagnetic waves are quantum mechanical: 
  they are boson fields. 
Currently, noises of quantum-mechanical origin limit the performance of advanced
  high-sensitivity photodetection systems
  \cite{sinclair19nanowire,mccaughan19nanowire,lee18tes}.
Therefore, analysis of the ultimate limits of any communications system requires
  quantum information theory \cite{wilde16quantumit2ed}.
This led to the development of the SRL for covert communication 
  over the lossy thermal noise bosonic channel, which is the underlying 
  quantum-mechanical description of many practical channels, including optical,
  microwave, and radio-frequency (RF) \cite{bash15covertbosoniccomm}.
\begin{figure}
	\centering
	\includegraphics[width=0.28\textwidth]{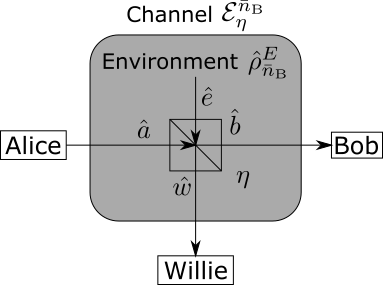}
	\caption{Single-mode bosonic channel $\mathcal{E}_{\eta}^{\bar{n}_{\rm B}}$ 
	modeled by a beamsplitter with transmissivity $\eta$ and an environment 
	injecting a thermal state $\hat{\rho}_{\bar{n}_{\rm B}}$ with mean photon
	number $\bar{n}_{\rm B}$. $\hat{a}$, $\hat{e}$, $\hat{b}$, and $\hat{w}$ 
	label input/output modal annihilation operators.}
	\label{fig:optchannel}
\end{figure}
The single-mode bosonic channel, depicted in Fig.~\ref{fig:optchannel} and formally
  defined in Section \ref{sec:channel_model}, is parametrized by the power 
  coupling (transmissivity) $\eta$ between the transmitter Alice and the intended
  receiver Bob, and the mean photon number $\bar{n}_{\rm B}$ per mode injected by
  the environment, where a single spatial-temporal-polarization mode is our
  fundamental transmission unit.
In our analysis (as in \cite{bash15covertbosoniccomm}) we do not assume a specific
  receiver structure for the adversary Willie.
Willie has access to all transmitted photons that are not captured
  by Bob, on which he can perform arbitrary quantum information 
  processing, including joint detection measurement, and use of unlimited quantum 
  memory and computing resources.
This makes our system \emph{quantum-secure}.
Furthermore, we assume that Willie has knowledge of all communication system 
  details (including the start time, center frequency, duration, and bandwidth of
  the transmission), except for a secret shared between Alice and Bob 
  before communication begins.
We use this secret to enable covertness irrespective of channel 
  conditions\footnote{While this assumption seems onerous, in many scenarios the 
  cost of having transmission detected greatly exceeds that of sharing a secret.
  Furthermore, classical results \cite{che13sqrtlawbscisit,bloch15covert} suggest 
  that the secret is unnecessary if Alice has a better channel to
  Bob than to Willie, however, ensuring this in practice may be harder than
  exchanging a secret.}
  and note that this meets the ``best practices'' of secure system design
  as the security of the system only depends on the shared secret
  \cite{menezes96HAC}. 
Finally, we assume existence of noise that is not under Willie's control.
Not only is this well-grounded, but also is necessary for covertness, as
  otherwise, transmissions cannot be hidden \cite[Th.~1]{bash15covertbosoniccomm}.
We use standard asymptotic notation \cite[Ch.~3.1]{clrs2e}, where 
  $f(n)=\mathcal{O}(g(n))$ and $f(n)=o(h(n))$ denote $g(n)$ and $h(n)$ as
  asymptotically tight and loose upper bounds on $f(n)$, respectively.
The SRL implies that the number $M$ of reliably transmissible covert bits 
  using $n$ modes is: 
\begin{align}
\label{eq:M}M&= \sqrt{n}\delta c_{\rm cov}c_{\rm rel}+o(\sqrt{n}),
\end{align}
where $\delta$ parametrizes the desired level of covertness (formally defined
  in Section \ref{sec:hyptest}),
  $c_{\rm cov}$ characterizes the mean transmitted photon number per mode 
  $\bar{n}_{\rm S}=\delta c_{\rm cov}/\sqrt{n}$ that is \emph{covert} given both 
  the channel and the modulation scheme, while $c_{\rm rel}$ captures the amount of
  information that can be transmitted \emph{reliably} (i.e., with arbitrarily small
  decoding error probability) by encoding it in $\bar{n}_{\rm S}$ photons/mode.
Our main focus is $c_{\rm cov}$, which determines the number of 
  covertly-transmissible photons.
We show that the optimal $c_{\rm cov}$ is:
\begin{align}
\label{eq:c_det}c_{\rm cov}&=\frac{\sqrt{2\eta\bar{n}_{\rm B}(1+\eta\bar{n}_{\rm B})}}{1-\eta},
\end{align}
and note that $c_{\rm cov}$ does not depend on Bob's receiver.
We then prove that it is achievable using quadrature phase shift keying (QPSK)
  modulation over coherent states (which describe ideal laser light 
  quantum-mechanically).
Since binary phase shift keying (BPSK) is known to achieve the Holevo capacity
  of (non-covert) communication over lossy thermal noise bosonic channel
  in the low received signal-to-noise ratio (SNR) regime
  \cite{lacerda17cohstateconstellations}, we evaluate its covertness properties.
We find that it is strictly suboptimal to QPSK, which further underscores 
  the differences between covert and non-covert communications.
However, the optimality of QPSK modulation leads to exact
  characterization of the optimal coding strategy and $c_{\rm rel}$.
We show how QPSK is combined with any channel code while maintaining
  covertness and describe how optimal $c_{\rm rel}$ is achieved in expectation.
We also discuss a promising approach to solving the general coding
  problem for covert communications over bosonic channels, leaving the full
  treatment to future work.
The work presented in this paper allows construction of communications systems for 
  many practical channels (including optical, microwave, RF, and others) that are
  provably covert against the most powerful adversaries allowed by the laws of 
  physics.
As such, these systems are future-proof.
Our results also have far-reaching implications beyond covert communication.
At the heart of our proof lies a new result on quantum state discrimination of
  a discrete set of displaced thermal states, which would lead to fundamental
  insights into optical state discrimination in loss and noise.
This has applications to optical communications and sensing, as well as structured
  designs for optimal receivers for these tasks---a topic wide open for
  future research.
This paper is organized as follows: next we present formally the
  lossy thermal noise bosonic channel model and the mathematical criteria for 
  covertness.
In Section \ref{sec:gaussian} we prove the converse by showing that our
  covertness criterion does not allow $c_{\rm cov}$ to exceed the right hand side
  (RHS) of \eqref{eq:c_det}.
In Section \ref{sec:discrete} we investigate discrete coherent state constellations,
  focusing on QPSK and BPSK, and show that QPSK achieves the RHS of \eqref{eq:c_det}
  while BPSK does not.
In Section \ref{sec:coding} we discuss the characterization of $c_{\rm rel}$ and
  the coding strategies for covert communication.
\section{Prerequisites}
\subsection{Channel model}
\label{sec:channel_model}
Consider a single mode lossy thermal noise channel ${\cal E}_{\eta}^{\bar{n}_{\rm B}}$ in Fig.~\ref{fig:optchannel}.
This is the quantum mechanical description of the transmission of a single 
  (spatio-temporal-polarization) mode of the electromagnetic field at a given
  transmission wavelength (such as optical or microwave) over linear loss and
  additive Gaussian noise (such as noise stemming from blackbody radiation).
A beamsplitter with transmissivity (fractional power coupling) $\eta$ models loss.
The input-output relationship between the bosonic mode operators of the single-mode 
  Alice-to-Bob channel, ${\hat b} = \sqrt{\eta}{\hat a} + {\sqrt{1-\eta}}{\hat e}$,  
  requires the ``environment'' mode ${\hat e}$ to ensure 
  $\left[ {\hat b}, {\hat b}^\dagger \right]=1$, and preserve the Heisenberg
  uncertainty law of quantum mechanics. 
Contrarily, power attenuation in a classical channel is captured by the relationship 
  $b = {\sqrt{\eta}} a$, where $a$ and $b$ are complex amplitudes of input and
  output mode functions. 
Bob captures a fraction $\eta$ of Alice's transmitted photons, while Willie is assumed to have access to the remaining $1-\eta$ fraction.
Noise is modeled by mode $\hat e$ being in a zero-mean thermal state 
  $\hat{\rho}_{\bar{n}_{\rm B}}$, which is expressed in the coherent state and Fock
  (photon number) bases as follows:
\begin{align}
\label{eq:thermal}\hat{\rho}_{\bar{n}_{\rm B}}&=\frac{1}{\pi \bar{n}_{\rm B}}\int_{\mathbb{C}}\exp\left[-\frac{|\alpha|^2}{\bar{n}_{\rm B}}\right]\dif^2\alpha\ket{\alpha}\bra{\alpha}=\sum_{k=0}^\infty t_k\ket{k}\bra{k},
\end{align}
where 
\begin{align}
\label{eq:tn}t_k=\frac{\bar{n}_{\rm B}^k}{(1+\bar{n}_{\rm B})^{k+1}}
\end{align}  
and $\bar{n}_{\rm B}$ is the mean photon number per mode injected by 
  the environment.
\begin{figure}
	\centering
	\includegraphics[width=0.95\textwidth]{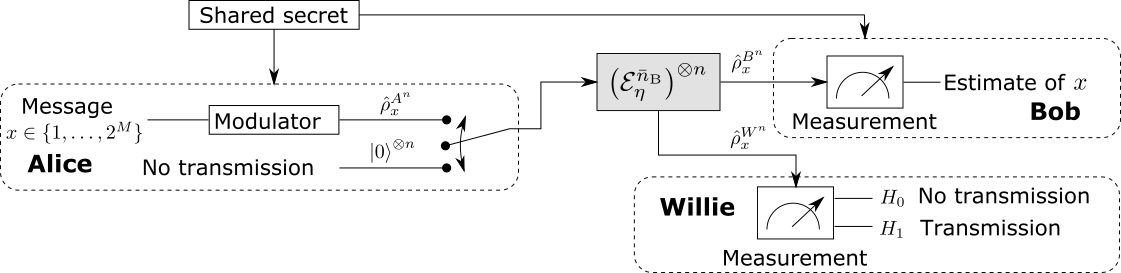}
	\caption{Covert communication over lossy thermal noise bosonic channel.
	Alice has a lossy thermal noise bosonic channel depicted in 
  Fig.~\ref{fig:optchannel} to legitimate receiver Bob and adversary Willie. 
  Alice encodes message $x$ with blocklength $n$ code, and chooses whether to
  transmit it using $\mathcal{E}_{\eta}^{\bar{n}_{\rm B}}$ $n$ times.
  Willie observes his channel from Alice to determine whether she is quiet 
  (null hypothesis $H_0$) or not (alternate hypothesis $H_1$). 
  Covert communication system must ensure that any detector Willie uses is close to 
  ineffective (i.e., a random guess between the hypotheses), while allowing Bob to 
  reliably decode the message (if one is transmitted). 
  Alice and Bob share a secret prior to transmission.}
	\label{fig:setup}
\end{figure}
Our covert communication framework is depicted in Fig.~\ref{fig:setup}.
We treat each mode as the fundamental transmission unit and assume that
  $n=2TW$ modes are available to Alice and Bob.
$TW$ is the number of orthogonal temporal modes, which is the product of 
  the transmission duration $T$ (in seconds) and the optical
  bandwidth $W$ (in Hz) of the source around its center frequency.
The factor of two corresponds to the use of both orthogonal polarizations.
Alice attempts to communicate reliably to Bob without 
  detection by Willie as depicted.
She uses a secret shared with Bob prior to the start of communication.
If she decides to transmit message $x$, she modulates an $n$-mode state 
  $\hat{\rho}^{A^n}_x$ using the shared secret.
While we assume that the bosonic channel acts on each input mode independently,
  $\hat{\rho}^{A^n}_x$ may be entangled across $n$ modes. 
Alice and Bob desire \emph{reliability}: for any $\epsilon>0$, Bob's decoding error
  probability $P^{\mathrm{(b)}}_{\rm e}\leq\epsilon$ for $n$ sufficiently large.
Bob may employ joint detection (entangling) measurement across $n$ modes.
Willie performs a quantum-optimal hypothesis test to determine whether Alice
  transmitted, which we discuss next.
\subsection{Hypothesis testing and covertness criteria}
\label{sec:hyptest}
As described in Fig.~\ref{fig:setup}, Willie observes a product thermal state 
  $\hat{\rho}^{W^n}_0=\hat{\rho}_{\eta\bar{n}_{\rm B}}^{\otimes n}$
  when Alice does not transmit and some other state $\hat{\rho}^{W^n}_1$ when
  she does.
Hypothesis $H_0$ corresponds to no transmission, and $H_1$ to transmission. 
Willie can err in raising a false alarm or missing Alice's transmission.
We denote Willie's probability of false alarm by 
  $P_{\mathrm{FA}}=P(\text{choose~}H_1|H_0)$ and his probability of missed 
  detection by
  $P_{\mathrm{MD}}=P(\text{choose~}H_0|H_1)$.
Assuming equally likely hypotheses $P(H_0)=P(H_1)=\frac{1}{2}$,
  Willie's detection error probability is
  $P^{\mathrm{(w)}}_{\rm e}=\frac{P_{\mathrm{FA}}+P_{\mathrm{MD}}}{2}$, which
  gives rise to the following covertness criterion:
\begin{criterion}
\label{crit:covcritP}
A system is covert if, for any $\delta_{\rm P}>0$, 
  $P^{\mathrm{(w)}}_{\rm e}\geq\frac{1}{2}-\delta_{\rm P}$ for $n$ large enough.
\end{criterion}
Subscript ``P'' refers to ``probability of detection error'' limit:
  since random decision results in $P^{\mathrm{(w)}}_{\rm e}=\frac{1}{2}$, small 
  $\delta_{\rm P}$ ensures that any detector that Willie constructs is similarly 
  ineffective.
Criterion \ref{crit:covcritP} applies even
  when the hypotheses are not equally likely \cite{sobers17jammer}.
A quantum-optimal receiver yields 
  $\min P^{\rm (w)}_{\rm e}=\frac{1}{2}-\frac{1}{4}\|\hat{\rho}^{W^n}_0-\hat{\rho}^{W^n}_1\|_1$,
  where $\|\hat{\rho}-\hat{\sigma}\|_1$
  is the trace distance between quantum states $\hat{\rho}$ and $\hat{\sigma}$
  \cite[Section 9.1.4]{wilde16quantumit2ed}.
Thus, Criterion \ref{crit:covcritP} is satisfied if 
  $\frac{1}{4}\|\hat{\rho}^{W^n}_0-\hat{\rho}^{W^n}_1\|_1\leq\delta_{\rm P}$.
However, quantum relative entropy (QRE) $D(\hat{\rho}\|\hat{\sigma})=\trace\left[\hat{\rho}\log\hat{\rho}-\hat{\rho}\log\hat{\sigma}\right]$
is a more convenient measure of covertness because it is additive over product
  states: 
$D\left(\hat{\rho}_1\otimes\hat{\rho}_2\|\hat{\sigma}_1\otimes\hat{\sigma}_2\right)=D\left(\hat{\rho}_1\|\hat{\sigma}_1\right)+D\left(\hat{\rho}_2\|\hat{\sigma}_2\right)$.
It is related to performance of optimal hypothesis test by the quantum 
  Chernoff-Stein lemma \cite{ogawa00stein} and
  Pinsker's inequality $\|\hat{\rho}-\hat{\sigma}\|_1\leq\sqrt{2D(\hat{\rho}\|\hat{\sigma})}$ \cite[Th.~10.8.1]{wilde16quantumit2ed}.
Therefore, instead of Criterion \ref{crit:covcritP},
  in the analysis that follows we use the following:
\begin{criterion}\label{crit:covcritRE}
A system is covert if, for any $\delta_{\rm QRE}>0$, 
  $D\left(\hat{\rho}^{W^n}_1\|\hat{\rho}^{W^n}_0\right)\leq\delta_{\mathrm{QRE}}$ for $n$ large enough.
\end{criterion}
By Pinsker's inequality, setting $\delta_{\rm QRE}=2\delta_{\rm P}^2$, Alice
  maintains a slightly higher level of covertness.
Classical version of Criterion \ref{crit:covcritRE} has been used in covert 
  communication proofs over standard classical channels
  \cite{wang15covert,bloch15covert}; we follow the same methodology here,
  setting $\delta=\sqrt{\delta_{\rm QRE}}$ in \eqref{eq:M}.
\section{Ultimate limit of covert communication over bosonic channel}
\label{sec:gaussian}
Criterion \ref{crit:covcritRE} imposes a constraint on Alice's transmitted mean 
  photon number per mode $\bar{n}_{\rm S}$:
\begin{theorem}[Converse]
\label{th:converse}
$D\left(\hat{\rho}^{W^n}_1\|\hat{\rho}^{W^n}_0\right)\leq\delta_{\mathrm{RE}}$ implies that $\bar{n}_{\rm S}\leq  \frac{\sqrt{2\eta\bar{n}_{\rm B}(1+\eta\bar{n}_{\rm B})}}{1-\eta}\sqrt{\frac{\delta_{\rm QRE}}{n}}$.
\end{theorem}
\begin{IEEEproof}
Alice transmits one of $2^M$ equally-likely $M$-bit messages by choosing an element
  from an arbitrary codebook  $\mathcal{C}=\{\hat{\rho}^{A^n}_x,x=1,\ldots,2^M\}$,
  where a state $\hat{\rho}^{A^n}_x=\ket{\psi_x}^{A^nA^n}\hspace{-4pt}\bra{\psi_x}$
  encodes an $M$-bit message $x$, and $\mathcal{C}$ is kept secret from Willie.
$\ket{\psi_x}^{A^n}=\sum_{\mathbf{m}\in\mathbb{N}_{0}^n}a_{\mathbf{m}}(x)\ket{\mathbf{m}}$ 
  is a general $n$-mode pure state, where
  $\ket{\mathbf{m}}\equiv\ket{m_1}\otimes\ket{m_2}\otimes\cdots\otimes\ket{m_n}$
  is a tensor product of $n$ Fock states. 
The mean photon number of an $n$-mode codeword $\hat{\rho}^{A^n}_x$ is
  $\bar{N}_{\mathrm{S},n}(x)=\sum_{\mathbf{m}\in\mathbb{N}_{0}^n}(\sum_{k=1}^nm_i)|a_{\mathbf{m}}(x)|^2$.
We limit our analysis to pure input states since, by convexity, using mixed
  states as inputs can only deteriorate the performance (it is
  equivalent to transmitting a randomly chosen pure state from an ensemble and 
  discarding the knowledge of that choice). 
When Alice transmits $\hat{\rho}^{A^n}_x$, Willie receives  
  a mixed state $\hat{\rho}^{W^n}_x$
  with the mean photon number 
  $(1-\eta)\bar{N}_{\mathrm{S},n}(x)+\eta n \bar{n}_{\rm B}$.
Willie does not have the codebook and must run a hypothesis test between 
  a product thermal state 
  $\hat{\rho}^{W^n}_0=\hat{\rho}_{\eta\bar{n}_{\rm B}}^{\otimes n}$ and a mixed state 
  $\bar{\rho}^{W^n}_1 = \frac{1}{2^M}\sum_{x=1}^{2^M} \hat{\rho}^{W^n}_x$.
The QRE is:
\begin{align}
\label{eq:entropyout}D\left(\bar{\rho}^{W^n}_1 \middle\| \hat{\rho}_{\eta\bar{n}_{\rm B}}^{\otimes n}\right) &= -S\left(\bar{\rho}^{W^n}_1\right)-\trace\left[\bar{\rho}^{W^n}_1\log\hat{\rho}_{\eta\bar{n}_{\rm B}}^{\otimes n}\right],
\end{align}
where $S(\rho)=-\trace[\rho \log \rho]$ is the von Neumann entropy.
Denote Willie's photon number operator associated with 
  the $k^{\text{th}}$ mode by $\hat{N}_k=\hat{w}_k^\dagger\hat{w}_k$, where
  $\hat{w}_k$ is Willie's annihilation operator associated with 
  the $k^{\text{th}}$ mode.
Since $\hat{N}_k$ is diagonal in Fock basis, by the properties of operator
  exponential,
\begin{align}
\label{eq:photonnumberpow}\hat{\rho}_{\eta\bar{n}_{\rm B}}^{\otimes n}&=\bigotimes_{k=1}^n\frac{1}{\eta\bar{n}_{\rm B}+1}\left(\frac{\eta\bar{n}_{\rm B}}{\eta\bar{n}_{\rm B}+1}\right)^{\hat{N}_k}
\end{align}
Substitution of \eqref{eq:photonnumberpow} into \eqref{eq:entropyout} yields:
\begin{align}
D\left(\bar{\rho}^{W^n}_1 \middle\| \hat{\rho}_{\eta\bar{n}_{\rm B}}^{\otimes n}\right) &=-S\left(\bar{\rho}^{W^n}_1\right)-\trace\left[\bar{\rho}^{W^n}_1\log \bigotimes_{k=1}^n\frac{1}{\eta\bar{n}_{\rm B}+1}\left(\frac{\eta\bar{n}_{\rm B}}{\eta\bar{n}_{\rm B}+1}\right)^{\hat{N}_k}\right]\\
&=-S\left(\bar{\rho}^{W^n}_1\right)-\trace\left[\bar{\rho}^{W^n}_1\sum_{k=1}^n\log \left[\frac{1}{\eta\bar{n}_{\rm B}+1}\left(\frac{\eta\bar{n}_{\rm B}}{\eta\bar{n}_{\rm B}+1}\right)^{\hat{N}_k}\right]\right]\\
&=-S\left(\bar{\rho}^{W^n}_1\right)-\trace\left[\bar{\rho}^{W^n}_1n\log \left[\frac{1}{\eta\bar{n}_{\rm B}+1}\right]+\log\left[\frac{\eta\bar{n}_{\rm B}}{\eta\bar{n}_{\rm B}+1}\right]\sum_{k=1}^n\bar{\rho}^{W^n}_1\hat{N}_k\right]\\
\label{eq:initkldivlb}&=-S\left(\bar{\rho}^{W^n}_1\right)+n\log[\eta\bar{n}_{\rm B}+1]-n[(1-\eta)\bar{n}_{\rm S}+\eta \bar{n}_{\rm B}]\log\left[\frac{\eta\bar{n}_{\rm B}}{\eta\bar{n}_{\rm B}+1}\right],
\end{align}
where \eqref{eq:initkldivlb} is because 
  $\bar{n}_{\rm S}=\frac{1}{n2^M}\sum_{x=1}^{2^M} \bar{N}_{\mathrm{S},n}(x)$.
Now, denote by $\bar{\rho}^{W}_{1,k}$ the state of the $k^{\text{th}}$ mode of 
  $\bar{\rho}^{W^n}_1$ that is obtained by tracing out the $n-1$ other modes.  
Let $\bar{n}_k$ be the mean photon number of $\bar{\rho}^{W}_{1,k}$.  
We upper-bound $S\left(\bar{\rho}^{W^n}_1\right)$ by:
\begin{align}
\label{eq:concavity}S\left(\bar{\rho}^{W^n}_1\right)&\overset{\text{(a)}}{\leq}\sum_{k=1}^nS\left(\bar{\rho}^{W^n}_{1,k}\right)\overset{\text{(b)}}{\leq}\sum_{k=1}^ng(\bar{n}_k)\overset{\text{(c)}}{\leq} ng((1-\eta)\bar{n}_{\rm S}+\eta \bar{n}_{\rm B}),
\end{align}
where (a) follows from the sub-additivity of the von Neumann entropy, (b) is 
  because the maximum von Neumann entropy of a single-mode state with mean photon 
  number constraint $\bar{n}$ is $g(\bar{n})$, where 
  $g(x)=(1+x)\log_2(1+x)-x\log_2x$~\cite{giovannetti04cappureloss}, and (c) follows
  from Jensen's inequality. 
Substituting \eqref{eq:concavity} into \eqref{eq:initkldivlb}, expanding $g(x)$, and
  re-arranging terms yields:
\begin{align}
D\left(\bar{\rho}^{W^n}_1 \middle\| \hat{\rho}_{\eta\bar{n}_{\rm B}}^{\otimes n}\right) &\geq n\left(((1-\eta)\bar{n}_{\rm S}+\eta \bar{n}_{\rm B})\log\left[1+\frac{(1-\eta)\bar{n}_{\rm S}}{\eta\bar{n}_{\rm B}}\right]\right.\nonumber\\
&\phantom{\geq n\bigg(\bigg.}-\left.(1+(1-\eta)\bar{n}_{\rm S}+\eta \bar{n}_{\rm B})\log\left[1+\frac{(1-\eta)\bar{n}_{\rm S}}{1+\eta\bar{n}_{\rm B}}\right]\right).
\end{align}
Since $x-\frac{x^2}{2}\leq\log(1+x)\leq x-\frac{x^2}{2}+\frac{x^3}{3}$
  for $x\geq 0$, we obtain:
\begin{align}
\label{eq:kl_ub}D\left(\bar{\rho}^{W^n}_1 \middle\| \hat{\rho}_{\eta\bar{n}_{\rm B}}^{\otimes n}\right) &\geq \frac{n(1-\eta)^2\bar{n}_{\rm S}^2}{2\eta\bar{n}_{\rm B}(1+\eta\bar{n}_{\rm B})}+o(\bar{n}_{\rm S}^2).
\end{align}
Discarding low-order terms, and solving \eqref{eq:kl_ub} for $\bar{n}_{\rm S}$
  yields the proof.
\end{IEEEproof}
The equality \eqref{eq:c_det} is implied by matching upper and lower bounds
  on $\bar{n}_{\rm S}$ in Theorem \ref{th:converse} and 
  \cite[Th.~2]{bash15covertbosoniccomm}, respectively.
However, the lower bound in \cite[Th.~2]{bash15covertbosoniccomm} is developed 
  from a random coding argument which uses an isotropic complex-valued
  Gaussian modulation of coherent states.
While such arguments are useful in mathematical proofs, they are a poor choice in 
  practice because of 1) exponential complexity of random codes, 2) unbounded
  storage required for complex numbers, and, 3) lack of peak power constraint.
Discrete modulation of coherent states is not only practical, but also
  achieves the Holevo capacity for the low 
  received SNR \cite{lacerda17cohstateconstellations}.
Discrete constellations also simplify coding: a polar code can be used
  over a discrete alphabet to achieve the channel capacity afforded by that
  alphabet.
Since covert communication naturally operates in the low SNR regime, we consider 
  the discrete modulation of coherent states next.
\section{Discrete modulation for covert communication over bosonic channels} \label{sec:discrete}
\subsection{Construction of transmitted sequence}
\label{sec:discrete_coding}
\begin{figure} 
    \centering
  \subfloat[Quadrature Phase Shift Keying (QPSK)\label{fig:qpsk}]{
  \hfill
       \includegraphics[width=0.45\linewidth]{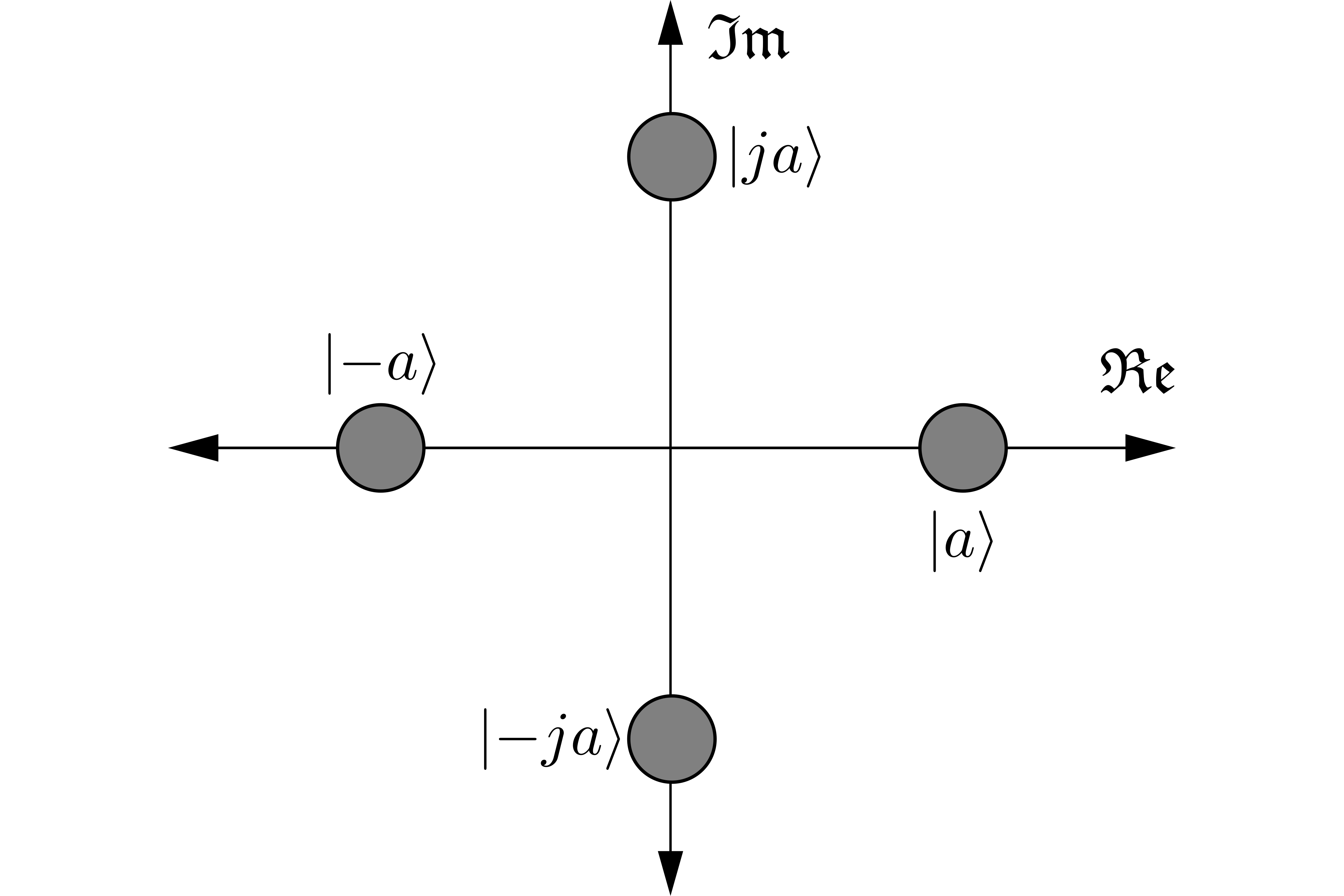}}
    \hfill
  \subfloat[Binary Phase Shift Keying (BPSK)\label{fig:bpsk}]{
        \includegraphics[width=0.45\linewidth]{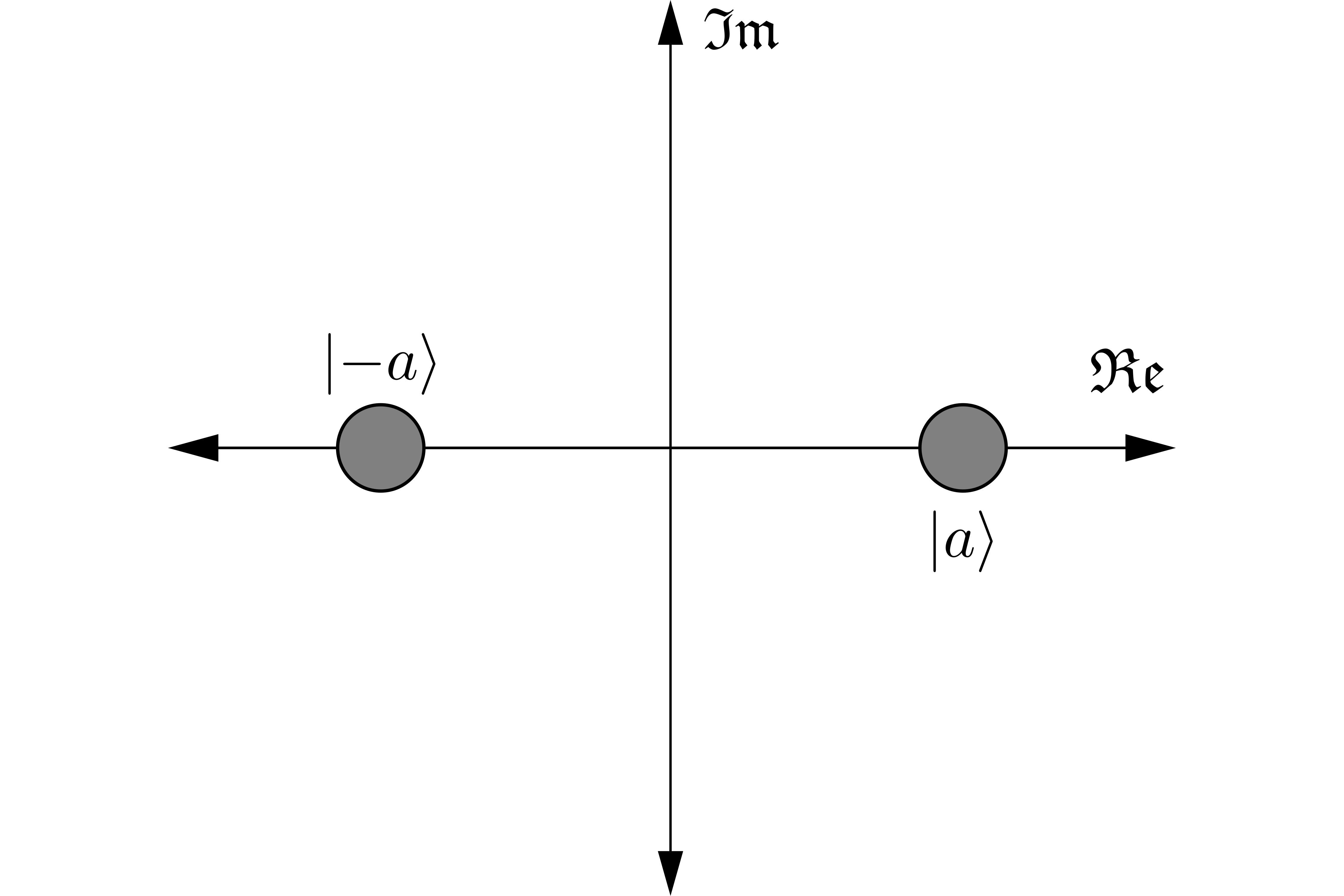}}
    \hfill
  \caption{Discrete coherent state constellations used by Alice's modulator.}
  \label{fig:psk} 
\end{figure}

Consider Alice transmitting an independent and identically distributed (i.i.d.)
  sequence $\mathbf{a}$ of $n$ symbols drawn from a discrete alphabet 
  $\mathcal{A}=\{a_l,l=1,\ldots,L,a_l\in\mathbb{C}\}$ with probability $p(l)$.
This corresponds a transmission using either:
\begin{itemize}
\item \textbf{Secret random code}: Alice and Bob secretly create a random code
  that maps $M$-bit input blocks to $n$-symbol codewords from
  $\mathcal{A}^n$ by
  generating $2^M$ codeword sequences
  $\mathcal{C}=\{\mathbf{c}(x)\}$, $x=1,\ldots, 2^M$ for messages $\{x\}$ 
  according to $p(\mathbf{c})=\prod_{k=1}^np(c_k)$ where 
  $p(c_k=a_l)=p(l)$.
\item \textbf{Secret random sequence}: Before communicating Alice and Bob 
  secretly draw a sequence $\mathbf{r}\in\{1,\ldots,L\}^n$ of length $n$
  where 
  $p(\mathbf{r})=\prod_{k=1}^np(r_k)$.
  Message $x$ is mapped to an $n$-symbol codeword 
  $\mathbf{c}(x)\in\{1,\ldots,L\}^n$ using a code that is known to Willie.
  Alice transmits a sequence from $\mathcal{A}^n$ corresponding to
  $\mathbf{a}=(\mathbf{c}(x)+\mathbf{r})\bmod L$, with element-wise modulo.
  Bob uses $\mathbf{r}$ to decode (e.g., by adding $\mathbf{r}$ modulo $L$ to 
  the received transmission before decoding).
\end{itemize}
We consider binary and quadrature shift keying (BPSK and QPSK) modulation 
  with corresponding alphabets $\mathcal{A}_{\rm b}=\{a,-ja\}$ and
  $\mathcal{A}_{\rm q}=\{a,ja,-a,-ja\}$.
Probabilities are $p(a)=p(-a)=\frac{1}{2}$ for BPSK and
  $p(a)=p(ja)=p(-a)=p(-ja)=\frac{1}{4}$ for QPSK.
On each mode, Alice transmits a coherent state with amplitudes corresponding to
  symbols from either $\mathcal{A}_{\rm b}$ or $\mathcal{A}_{\rm q}$, with
  the resulting constellations depicted in Fig.~\ref{fig:psk}.
\subsection{Willie's received state}
When Alice transmits $\ket{a}$, Willie receives a
  displaced thermal state 
  $\hat{\rho}_{\eta\bar{n}_{\rm B}}(\sqrt{1-\eta}a)$, where
\begin{align}
\label{eq:rhonbB}\hat{\rho}_{\bar{n}_{\rm B}}(a)&=\frac{1}{\pi \bar{n}_{\rm B}}\int_{\mathbb{C}}\exp\left[-\frac{|\beta-a|^2}{\bar{n}_{\rm B}}\right]\dif^2\beta\ket{\beta}\bra{\beta}.
\end{align}
However, Alice's scheme described in Section \ref{sec:discrete_coding}
  results in Willie observing a mixture 
  $\hat{\rho}^W_{1,L}=\sum_{l=1}^L p(l)\hat{\rho}_{\eta\bar{n}_{\rm B}}(\sqrt{1-\eta}a_l)$
  of $L$ displaced thermal states in each of $n$ modes.
This is because the secret random code has no structure of use to Willie, and 
  the secret random 
  sequence\footnote{This scheme resembles an application of a one-time pad that is
  typically used in cryptography to ensure absolute secrecy \cite{shannon49sec}. 
  Its use for covert communication is described in 
  \cite[Remark after Th.~1.2]{bash13squarerootjsacnonote}.}
  destroys any structure in the public code that could be used by Willie. 
Note that neither the transmitted codeword from the random codebook nor the random 
  sequence $\mathbf{r}$ can be re-used.
Since Alice's modulated sequence is i.i.d., Willie observes
  $\hat{\rho}^{W^n}_{1,L}=\left(\hat{\rho}^W_{1,L}\right)^{\otimes n}$.
The QRE is thus:
\begin{align}
D\left(\hat{\rho}^{W^n}_{1,L}\|\hat{\rho}^{W^n}_0\right)&=nD\left(\hat{\rho}^W_{1,L}\|\hat{\rho}_{\eta \bar{n}_{\rm B}}\right).
\end{align}
Thus, to maintain Criterion \ref{crit:covcritRE}, Alice must employ a modulation
  scheme such that
\begin{align}
\label{eq:qreineq}D\left(\hat{\rho}^W_{1,L}\|\hat{\rho}_{\eta \bar{n}_{\rm B}}\right)\leq\frac{\delta_{\rm QRE}}{n}.
\end{align} 
Next we prove that QPSK constellation achieves the fundamental limit of covert communication over the lossy thermal noise bosonic channel. This is because it allows the transmission of the maximal mean photon number characterized by \eqref{eq:c_det} while maintaining Criterion \ref{crit:covcritRE}. 
Since BPSK modulation achieves the Holevo capacity for the low 
  received SNR regime \cite{lacerda17cohstateconstellations}, which
  is natural for covert communication, we analyze the performance of BPSK.
We show that it is strictly suboptimal, and that maintaining covertness requires
  reducing the mean photon number over QPSK, further underscoring the differences
   between covert and non-covert communications.
We conclude this section by showing how to make a constant amplitude QPSK constellation covert, and allow the use of practical channel codes.
Before we continue, we state useful lemmas.
Suppose $\hat{A}(t)$ and $\hat{B}(t)$ are non-singular operators parametrized by
  $t$, and $\hat{I}$ is the identity operator.
Then the following two lemmas hold:
\begin{lemma}[{\cite[Th.~6]{haber18matrixexplog}}]
\label{lemma:dlogint}
$\frac{\mathrm{d}}{\mathrm{d}t}\log \hat{A}(t)=\int_0^1\mathrm{d}s\left[s\hat{A}(t)+(1-s)\hat{I}\right]^{-1}\frac{\mathrm{d}\hat{A}(t)}{\mathrm{d}t}\left[s\hat{A}(t)+(1-s)\hat{I}\right]^{-1}$.
\end{lemma}
\begin{lemma}[{\cite[lemma in Sec.~4]{haber18matrixexplog}}]
\label{lemma:dinv}
$\frac{\mathrm{d}}{\mathrm{d}t}B^{-1}(t)=-B^{-1}(t)\frac{\mathrm{d}\hat{B}(t)}{\mathrm{d}t}B^{-1}(t)$.
\end{lemma}
\renewcommand{\kh}{\hat{K}_{\QPSK}} 
\renewcommand{\rz}{\hat{\rho}_{1,\QPSK}} 
\newcommand{\sumvar}{k}
\renewcommand{\nt}{\bar{n}_{\rm T}}
\subsection{Quadrature phase shift keying}\label{sec:QPSK}
\label{sec:optimalQPSK}
\begin{theorem}
\label{th:optimalQPSK}
QPSK modulation achieves $\bar{n}_{\rm S}\leq  \frac{\sqrt{2\eta\bar{n}_{\rm B}(1+\eta\bar{n}_{\rm B})}}{1-\eta}\sqrt{\frac{\delta_{\rm QRE}}{n}}$ while maintaining\\ $D\left(\hat{\rho}^{W^n}_1\|\hat{\rho}^{W^n}_0\right)=nD\left(\hat{\rho}^W_{1,m}\|\hat{\rho}_{\eta \bar{n}_{\rm B}}\right)$.
\end{theorem}
\begin{IEEEproof}
Consider $\rz=\frac{1}{4}\left(\ra+\rb+\rc+\rd\right)$, as the equal-weighted mixture of displaced thermal states where $\ra \equiv \hat{\rho}_{\nt}(u)$, $\rb \equiv \hat{\rho}_{\nt}(ju)$, $\rc \equiv \hat{\rho}_{\nt}(-u)$, and $\rd \equiv \hat{\rho}_{\nt}(-ju)$ with $\hat{\rho}_{\nt}(\beta)$ defined in \eqref{eq:rhonbB}. 
Subscript ``q'' stands for QPSK, since setting $u=\sqrt{1-\eta}b$ and $\nt= \eta\bar{n}_{\rm B}$ yields $\ra$, $\rb$, $\rc$, and $\rd$ as the displaced thermal states observed by Willie when Alice transmits $\ket{b}$, $\ket{{\im}b}$, $\ket{-b}$, and $\ket{-{\im}b}$, respectively, and zero-mean thermal state $\hat{\rho}_0\equiv\hat{\rho}_{\nt}(0)$ when she does not transmit. 
Thus, setting $\hat{\rho}_{1,m}^{W} = \rz$ and dropping $W$ from superscript for brevity yields $D\left(\rz\|\ro\right)$ as the left hand side of \eqref{eq:qreineq}. There are no known closed form expressions for $D\left(\rz\|\ro\right)$, therefore, we evaluate its Taylor series expansion. 
To do so, we must find the first four derivatives of $\rz$ with respect to displacement $u$, and set $u=0$. The derivatives of $\ra$, $\rb$, $\rc$, and $\rd$ are as follows
  \cite[Ch.~VI, Eq.~(1.31)]{helstrom76quantumdetect}:
\begin{align}
\label{eq:drho00}\frac{\d\ra}{{\d}u}&=\nt^{-1}\left((\hat{a} - u)\ra\ + \ra(\hat{a}^\dagger - u)\right),\\
\label{eq:drho01}\frac{\d\rb}{{\d}u} &= -\nt^{-1}\left((j\hat{a} + u)\rb\ - \rb(j\hat{a}^\dagger - u)\right),\\
\label{eq:drho10}\frac{\d\rc}{{\d}u} &= -\nt^{-1}\left((\hat{a} + u)\rc\ + \rc(\hat{a}^\dagger + u)\right),\\
\label{eq:drho11}\frac{\d\rd}{{\d}u} &= \nt^{-1}\left((j\hat{a} - u)\rd\ - \rd(j\hat{a}^\dagger + u)\right),
\end{align}
where $\hat{a}^\dagger$ and $\hat{a}$ denote Alice's creation and annihilation operators,
  respectively.
These allow us to differentiate $\rz$ with respect to displacement $u$. For each, setting $u = 0$ yields:
\begin{align}
\label{eq:drho13u0}\frac{\d\rz}{{\d}u}\big\rvert_{u=0} &=\frac{\d^3\rz}{{\d}u^3}\big\rvert_{u=0} = 0,\\
\label{eq:drho2u0}\frac{\d^2\rz}{{\d}u^2}\big\rvert_{u=0} &= \frac{2}{\nt^2}\left(\ah\ro\ad\right)-\frac{2}{\nt}\left(\ro\right),\\
\label{eq:drho4u0}\frac{\d^4\rz}{{\d}u^4}\big\rvert_{u=0} &= \frac{12\ro}{\nt^2}-\frac{6}{\nt^3}(\ah\ro\ad)+\frac{1}{\nt^4}\left(\ah^4\ro+6\ah^2\ro(\ad)^2+\ro(\ad)^4\right)
\end{align} \par
Denote by $\kh = \rz\log\rz - \rz\log\ro$ the term inside the trace in the definition of QRE $D\left(\rz\|\ro\right)$. Now, let's evaluate each term of the Taylor series expansion of $D\left(\rz\|\ro\right)$. 
\subsubsection{First term}
Using Lemma \ref{lemma:dlogint}, the first derivative of $\kh$ with respect to $u$ 
  is as follows:
\begin{align}
\frac{\mathrm{d}\kh}{\mathrm{d}u}&=\frac{\mathrm{d}\rz}{\mathrm{d}u}\log\rz+\rz\int_0^1\mathrm{d}s\hat{\sigma}_1^{-1}(s)\frac{\mathrm{d}\rz}{\mathrm{d}u}\hat{\sigma}_1^{-1}(s)-\frac{\mathrm{d}\rz}{\mathrm{d}u}\log\hat{\rho}_0,
\end{align}
where $\hat{\sigma}_1(s)=s\rz+(1-s)\hat{I}$.
Setting $u=0$ yields:
\begin{align}
\left.\frac{\mathrm{d}\kh}{\mathrm{d}u}\right|_{u=0}&=\left.\frac{\mathrm{d}\rz}{\mathrm{d}u}\right|_{u=0}\log\hat{\rho}_0+\hat{\rho}_0\int_0^1\mathrm{d}s\hat{\sigma}_0^{-1}(s)\left.\frac{\mathrm{d}\rz}{\mathrm{d}u}\right|_{u=0}\hat{\sigma}_0^{-1}(s)-\left.\frac{\mathrm{d}\rz}{\mathrm{d}u}\right|_{u=0}\log\hat{\rho}_0\\
&=\hat{\rho}_0\int_0^1\mathrm{d}s\hat{\sigma}_0^{-1}(s)\left.\frac{\mathrm{d}\rz}{\mathrm{d}u}\right|_{u=0}\hat{\sigma}_0^{-1}(s),
\end{align}
where $\hat{\sigma}_0(s)=s\hat{\rho}_0+(1-s)\hat{I}$.
Since $\left.\frac{\mathrm{d}\rz}{\mathrm{d}u}\right|_{u=0}=0$ by \eqref{eq:drho13u0}, $\left.\frac{\mathrm{d}\kh}{\mathrm{d}u}\right|_{u=0}=0$.
Thus, $\trace\left[\frac{\mathrm{d}\kh}{\mathrm{d}u}\big|_{u=0}\right]$.
\subsubsection{Second term}
Using Lemma \ref{lemma:dinv}, the second derivative of $\kh$ with respect to $u$ 
  is as follows:
\begin{align}
\nonumber\frac{\mathrm{d}^2\kh}{\mathrm{d}u^2}&=2\frac{\mathrm{d}\rz}{\mathrm{d}u}\int_0^1\mathrm{d}s\hat{\sigma}_1^{-1}(s)\frac{\mathrm{d}\rz}{\mathrm{d}u}\hat{\sigma}_1^{-1}(s)-2\rz\int_0^1s\mathrm{d}s\hat{\sigma}_1^{-1}(s)\frac{\mathrm{d}\rz}{\mathrm{d}u}\hat{\sigma}_1^{-1}(s)\frac{\mathrm{d}\rz}{\mathrm{d}u}\hat{\sigma}_1^{-1}(s)\\
\label{eq:dK2QPSK}&\phantom{=}+\rz\int_0^1\mathrm{d}s\hat{\sigma}_1^{-1}(s)\frac{\mathrm{d}^2\rz}{\mathrm{d}u^2}\hat{\sigma}_1^{-1}(s)+\frac{\mathrm{d}^2\rz}{\mathrm{d}u^2}\log\rz-\frac{\mathrm{d}^2\rz}{\mathrm{d}u^2}\log\hat{\rho}_0.
\end{align}
Setting $u=0$ in \eqref{eq:dK2QPSK}, discarding terms containing 
  $\left.\frac{\mathrm{d}\rz}{\mathrm{d}u}\right|_{u=0}=0$, and
  canceling the positive and negative 
  $\left.\frac{\mathrm{d}^2\rz}{\mathrm{d}u^2}\right|_{u=0}\log\hat{\rho}_0$,
  we have:
\begin{align}
\label{eq:dK2u0Q}\left.\frac{\mathrm{d}^2\kh}{\mathrm{d}u^2}\right|_{u=0}&=\hat{\rho}_0\int_0^1\mathrm{d}s\hat{\sigma}_0^{-1}(s)\left.\frac{\mathrm{d}^2\rz}{\mathrm{d}u^2}\right|_{u=0}\hat{\sigma}_0^{-1}(s)
\end{align}
Substitution of \eqref{eq:drho2u0} into \eqref{eq:dK2u0Q} yields the following:
\begin{align}
\label{eq:dK2u0Qeval}\frac{\mathrm{d}^2\kh(u)}{\mathrm{d}u^2}\bigg\rvert_{u=0}&=\frac{2}{\nt^2}\hat{\rho}_0\int_0^1\mathrm{d}s\hat{\sigma}_0^{-1}(s)\hat{a}\hat{\rho}_0\hat{a}^\dagger\hat{\sigma}_0^{-1}(s)
- \frac{2}{\nt}\hat{\rho}_0\int_0^1\mathrm{d}s\hat{\sigma}_0^{-1}(s)\hat{\rho}_0\hat{\sigma}_0^{-1}(s)
\end{align}
Now note that $\hat{\sigma}_0(s)$ is diagonal in the Fock state basis, implying:
\begin{align}
\hat{\sigma}_0^{-1}(s)&=\sum_{\sumvar=0}^\infty(st_\sumvar+(1-s))^{-1}\ket{\sumvar}\bra{\sumvar},
\end{align}
where we implicitly substitute $\nt$ for $\bar{n}_{\rm B}$ in \eqref{eq:tn}.
Now,
\begin{align}
\label{eq:intidentityQ}\int_0^1\mathrm{d}s\hat{\sigma}_0^{-1}(s)\hat{\rho}_0\hat{\sigma}_0^{-1}(s)&=\int_0^1\mathrm{d}s\sum_{\sumvar=0}^\infty t_\sumvar(st_\sumvar+(1-s))^{-2}\ket{\sumvar}\bra{\sumvar}=\hat{I}\\
\nonumber\int_0^1\mathrm{d}s\hat{\sigma}_0^{-1}(s)\hat{a}\hat{\rho}_0\hat{a}^\dagger\hat{\sigma}_0^{-1}(s)&=\int_0^1\mathrm{d}s\sum_{\sumvar=0}^\infty(\sumvar+1)t_{\sumvar+1}(st_\sumvar+(1-s))^{-2}\ket{\sumvar}\bra{\sumvar}\\
\label{eq:intaadaggerQ}&=\frac{\nt}{1+\nt}\sum_{\sumvar=0}^\infty(\sumvar+1)\ket{\sumvar}\bra{n},
\end{align}
since $\int_0^1\mathrm{d}s(sq+(1-s))^{-2}=\frac{1}{q}$ for $q>0$. 
Here, the traces of the two terms in \eqref{eq:dK2u0Qeval} cancel.
Thus, 
  $\trace\left[\left.\frac{\mathrm{d}^2\kh}{\mathrm{d}u^2}\right|_{u=0}\right]=0$.
\subsubsection{Third term}
Again using Lemma \ref{lemma:dinv}, the third derivative of $\kh$ with respect to 
  $u$ is:
\begin{align}
\nonumber\frac{\mathrm{d}^3\kh}{\mathrm{d}u^3}&=3\frac{\mathrm{d}^2\rz}{\mathrm{d}u^2}\int_0^1\mathrm{d}s\hat{\sigma}_1^{-1}(s)\frac{\mathrm{d}\rz}{\mathrm{d}u}\hat{\sigma}_1^{-1}(s)-6\frac{\mathrm{d}\rz}{\mathrm{d}u}\int_0^1s\mathrm{d}s\hat{\sigma}_1^{-1}(s)\frac{\mathrm{d}\rz}{\mathrm{d}u}\hat{\sigma}_1^{-1}(s)\frac{\mathrm{d}\rz}{\mathrm{d}u}\hat{\sigma}_1^{-1}(s)\\
\nonumber&\phantom{=}+3\frac{\mathrm{d}\rz}{\mathrm{d}u}\int_0^1\mathrm{d}s\hat{\sigma}_1^{-1}(s)\frac{\mathrm{d}^2\rz}{\mathrm{d}u^2}\hat{\sigma}_1^{-1}(s)-3\rz\int_0^1s\mathrm{d}s\hat{\sigma}_1^{-1}(s)\frac{\mathrm{d}^2\rz}{\mathrm{d}u^2}\hat{\sigma}_1^{-1}(s)\frac{\mathrm{d}\rz}{\mathrm{d}u}\hat{\sigma}_1^{-1}(s)\\
\nonumber&\phantom{=}+6\rz\int_0^1s^2\mathrm{d}s\hat{\sigma}_1^{-1}(s)\frac{\mathrm{d}\rz}{\mathrm{d}u}\hat{\sigma}_1^{-1}(s)\frac{\mathrm{d}\rz}{\mathrm{d}u}\hat{\sigma}_1^{-1}(s)\frac{\mathrm{d}\rz}{\mathrm{d}u}\hat{\sigma}_1^{-1}(s)\\
\nonumber&\phantom{=}-3\rz\int_0^1s\mathrm{d}s\hat{\sigma}_1^{-1}(s)\frac{\mathrm{d}\rz}{\mathrm{d}u}\hat{\sigma}_1^{-1}(s)\frac{\mathrm{d}^2\rz}{\mathrm{d}u^2}\hat{\sigma}_1^{-1}(s)\\
\label{eq:dK3Q}&\phantom{=}+\rz\int_0^1\mathrm{d}s\hat{\sigma}_1^{-1}(s)\frac{\mathrm{d}^3\rz}{\mathrm{d}u^3}\hat{\sigma}_1^{-1}(s)+ \frac{\mathrm{d}^3\rz}{\mathrm{d}u^3}\log\rz-\frac{\mathrm{d}^3\rz}{\mathrm{d}u^3}\log\hat{\rho}_0.
\end{align}
Since $\left.\frac{\mathrm{d}\rz}{\mathrm{d}u}\right|_{u=0}=0$ and $\left.\frac{\mathrm{d}^3\rz}{\mathrm{d}u^3}\right|_{u=0}=0$  by \eqref{eq:drho13u0},
  $\left.\frac{\mathrm{d}^3\kh}{\mathrm{d}u^3}\right|_{u=0}=0$.
\subsubsection{Fourth term}
We use Lemma \ref{lemma:dinv} once again, however, for brevity we omit writing
  terms containing $\frac{\mathrm{d}\rz}{\mathrm{d}u}$ and
  $\frac{\mathrm{d}^3\rz}{\mathrm{d}u^3}$, as these are zero operators
  when $u=0$.
Therefore, we have:
\begin{align}
\nonumber\frac{\mathrm{d}^4\kh}{\mathrm{d} u^4}&=6\frac{\mathrm{d}^2\rz}{\mathrm{d} u^2}\int_0^1\mathrm{d}s\hat{\sigma}_1^{-1}(s)\frac{\mathrm{d}^2\rz}{\mathrm{d} u^2}\hat{\sigma}_1^{-1}(s)-6\rz\int_0^1s\mathrm{d}s\hat{\sigma}_1^{-1}(s)\frac{\mathrm{d}^2\rz}{\mathrm{d} u^2}\hat{\sigma}_1^{-1}(s)\frac{\mathrm{d}^2\rz}{\mathrm{d} u^2}\hat{\sigma}_1^{-1}(s)\\
\label{eq:dK4Q}&\phantom{=}+\rz\int_0^1\mathrm{d}s\hat{\sigma}_1^{-1}(s)\frac{\mathrm{d}^4\rz}{\mathrm{d} u^4}\hat{\sigma}_1^{-1}(s)+\frac{\mathrm{d}^4\rz}{\mathrm{d}^4 u}\log\rz-\frac{\mathrm{d}^4\rz}{\mathrm{d}^4 u}\log\hat{\rho}_0.
\end{align}
Setting $u=0$ yields:
\begin{align}
\nonumber\left.\frac{\mathrm{d}^4\kh(u)}{\mathrm{d} u^4}\right|_{u=0}&=6\left.\frac{\mathrm{d}^2\rz}{\mathrm{d} u^2}\right|_{u=0}\int_0^1\mathrm{d}s\hat{\sigma}_0^{-1}(s)\left.\frac{\mathrm{d}^2\rz}{\mathrm{d} u^2}\right|_{u=0}\hat{\sigma}_0^{-1}(s)\\
\nonumber&\phantom{=}-6\hat{\rho}_0\int_0^1s\mathrm{d}s\hat{\sigma}_0^{-1}(s)\left.\frac{\mathrm{d}^2\rz}{\mathrm{d} u^2}\right|_{u=0}\hat{\sigma}_0^{-1}(s)\left.\frac{\mathrm{d}^2\rz}{\mathrm{d} u^2}\right|_{u=0}\hat{\sigma}_0^{-1}(s)\\
\label{eq:dK4u0Q}&\phantom{=}+\hat{\rho}_0\int_0^1\mathrm{d}s\hat{\sigma}_0^{-1}(s)\left.\frac{\mathrm{d}^4\rz}{\mathrm{d} u^4}\right|_{u=0}\hat{\sigma}_0^{-1}(s).
\end{align}
Substitution of \eqref{eq:drho2u0} in the first term of \eqref{eq:dK4u0Q} and taking
  the trace yields:
\begin{IEEEeqnarray}{Lll}
	\IEEEeqnarraymulticol{3}{l}{
		\nonumber\trace\left[6\left.\frac{\mathrm{d}^2\rz}{\mathrm{d} u^2}\right|_{u=0}\int_0^1\mathrm{d}s\hat{\sigma}_0^{-1}(s)\left.\frac{\mathrm{d}^2\rz}{\mathrm{d} u^2}\right|_{u=0}\hat{\sigma}_0^{-1}(s)\right]
	}\nonumber\\ \qquad \qquad
	\nonumber & = & \frac{24}{\nt^4}\trace\left[\hat{a}\hat{\rho}_0\hat{a}^\dagger\int_0^1\mathrm{d}s\hat{\sigma}_0^{-1}(s)\hat{a}\hat{\rho}_0\hat{a}^\dagger\hat{\sigma}_0^{-1}(s)\right]-\frac{24}{\nt^3}\trace\left[\hat{a}\hat{\rho}_0\hat{a}^\dagger\int_0^1\mathrm{d}s\hat{\sigma}_0^{-1}(s)\hat{\rho}_0\hat{\sigma}_0^{-1}(s)\right]
	\\
	\label{eq:tr1dK4u0Q} & & -\frac{24}{\nt^3}\trace\left[\hat{\rho}_0\int_0^1\mathrm{d}s\hat{\sigma}_0^{-1}(s)\hat{a}\hat{\rho}_0\hat{a}^\dagger\hat{\sigma}_0^{-1}(s)\right]+\frac{24}{\nt^2}\trace\left[\hat{\rho}_0\int_0^1\mathrm{d}s\hat{\sigma}_0^{-1}(s)\hat{\rho}_0\hat{\sigma}_0^{-1}(s)\right].
\end{IEEEeqnarray}
The four terms in \eqref{eq:tr1dK4u0Q} are evaluated using 
  \eqref{eq:intidentityQ} and \eqref{eq:intaadaggerQ}:
\begin{align}
\label{eq:4thtermpt1}\frac{24}{\nt^4}\trace\left[\hat{a}\hat{\rho}_0\hat{a}^\dagger\int_0^1\mathrm{d}s\hat{\sigma}_0^{-1}(s)\hat{a}\hat{\rho}_0\hat{a}^\dagger\hat{\sigma}_0^{-1}(s)\right]&=\frac{24(1+2\nt)}{\nt^2(1+\nt)}\\
-\frac{24}{\nt^3}\trace\left[\hat{a}\hat{\rho}_0\hat{a}^\dagger\int_0^1\mathrm{d}s\hat{\sigma}_0^{-1}(s)\hat{\rho}_0\hat{\sigma}_0^{-1}(s)\right]&=-\frac{24}{\nt^2}\\
-\frac{24}{\nt^3}\trace\left[\hat{\rho}_0\int_0^1\mathrm{d}s\hat{\sigma}_0^{-1}(s)\hat{a}\hat{\rho}_0\hat{a}^\dagger\hat{\sigma}_0^{-1}(s)\right]&=-\frac{24}{\nt^2}\\
\label{eq:tr1dK4u0c6Q}\frac{24}{\nt^2}\trace\left[\hat{\rho}_0\int_0^1\mathrm{d}s\hat{\sigma}_0^{-1}(s)\hat{\rho}_0\hat{\sigma}_0^{-1}(s)\right]&=\frac{24}{\nt^2}.
\end{align}
Summing \eqref{eq:4thtermpt1}-\eqref{eq:tr1dK4u0c6Q} yields the first term of 
  \eqref{eq:dK4u0Q}:
\begin{align}
\label{eq:tr1dK4u0evalQ}\trace\left[6\left.\frac{\mathrm{d}^2\rz}{\mathrm{d} u^2}\right|_{u=0}\int_0^1\mathrm{d}s\hat{\sigma}_0^{-1}(s)\left.\frac{\mathrm{d}^2\rz}{\mathrm{d} u^2}\right|_{u=0}\hat{\sigma}_0^{-1}(s)\right]&=\frac{24(1+2\nt)}{\nt^2(1+\nt)}-\frac{24}{\nt^2}
\end{align}
Substitution of \eqref{eq:drho2u0} in the second term of \eqref{eq:dK4u0Q} and taking
  the trace yields:
\begin{IEEEeqnarray}{rCl}
	\IEEEeqnarraymulticol{3}{l}{
		\trace\left[-6\hat{\rho}_0\int_0^1s\mathrm{d}s\hat{\sigma}_0^{-1}(s)\left.\frac{\mathrm{d}^2\rz}{\mathrm{d} u^2}\right|_{u=0}\hat{\sigma}_0^{-1}(s)\left.\frac{\mathrm{d}^2\rz}{\mathrm{d} u^2}\right|_{u=0}\hat{\sigma}_0^{-1}(s)\right]
	}\nonumber\\* \qquad \qquad \qquad
	\nonumber & = & -\frac{24}{\nt^4}\trace\left[\hat{\rho}_0\int_0^1s\mathrm{d}s\hat{\sigma}_0^{-1}(s)\hat{a}\hat{\rho}_0\hat{a}^\dagger\hat{\sigma}_0^{-1}(s)\hat{a}\hat{\rho}_0\hat{a}^\dagger\hat{\sigma}_0^{-1}(s)\right]
	\\
	\nonumber & & +\frac{24}{\nt^3}\trace\left[\hat{\rho}_0\int_0^1s\mathrm{d}s\hat{\sigma}_0^{-1}(s)\hat{\rho}_0\hat{\sigma}_0^{-1}(s)\hat{a}\hat{\rho}_0\hat{a}^\dagger\hat{\sigma}_0^{-1}(s)\right]
	\\
	\nonumber & & +\frac{24}{\nt^3}\trace\left[\hat{\rho}_0\int_0^1s\mathrm{d}s\hat{\sigma}_0^{-1}(s)\hat{a}\hat{\rho}_0\hat{a}^\dagger\hat{\sigma}_0^{-1}(s)\hat{\rho}_0\hat{\sigma}_0^{-1}(s)\right]
	\\
	\label{eq:tr2dK4u0Q} & & -\frac{24}{\nt^2}\trace\left[\hat{\rho}_0\int_0^1s\mathrm{d}s\hat{\sigma}_0^{-1}(s)\hat{\rho}_0\hat{\sigma}_0^{-1}(s)\hat{\rho}_0\hat{\sigma}_0^{-1}(s)\right].
\end{IEEEeqnarray}
Since $\int_0^1s\mathrm{d}s(sq+(1-s))^{-3}=\frac{1}{2q^2}$ for $q>0$, 
\begin{align}
\nonumber\int_0^1s\mathrm{d}s\hat{\sigma}_0^{-1}(s)\hat{a}\hat{\rho}_0\hat{a}^\dagger\hat{\sigma}_0^{-1}(s)\hat{a}\hat{\rho}_0\hat{a}^\dagger\hat{\sigma}_0^{-1}(s)&=\int_0^1s\mathrm{d}s\sum_{\sumvar=0}^\infty\frac{(\sumvar+1)^2t_{\sumvar+1}^2\ket{\sumvar}\bra{\sumvar}}{(st_\sumvar+(1-s))^3}\\
&=\frac{\nt^2}{2(1+\nt)^2}\sum_{\sumvar=0}^\infty(\sumvar+1)^2\ket{\sumvar}\bra{\sumvar}.
\end{align}
Thus, the first term of \eqref{eq:tr2dK4u0Q} is:
\begin{align}
\label{eq:tr2dK4u0c3Q}-\frac{24}{\nt^4}\trace\left[\hat{\rho}_0\int_0^1s\mathrm{d}s\hat{\sigma}_0^{-1}(s)\hat{a}\hat{\rho}_0\hat{a}^\dagger\hat{\sigma}_0^{-1}(s)\hat{a}\hat{\rho}_0\hat{a}^\dagger\hat{\sigma}_0^{-1}(s)\right]&=-\frac{12(1+2\nt)}{\nt^2(1+\nt)}.
\end{align}
Using a similar approach, we obtain the other terms:
\begin{align}
\frac{24}{\nt^3}\trace\left[\hat{\rho}_0\int_0^1s\mathrm{d}s\hat{\sigma}_0^{-1}(s)\hat{\rho}_0\hat{\sigma}_0^{-1}(s)\hat{a}\hat{\rho}_0\hat{a}^\dagger\hat{\sigma}_0^{-1}(s)\right]&=\frac{12}{\nt^2}\\
\frac{24}{\nt^3}\trace\left[\hat{\rho}_0\int_0^1s\mathrm{d}s\hat{\sigma}_0^{-1}(s)\hat{a}\hat{\rho}_0\hat{a}^\dagger\hat{\sigma}_0^{-1}(s)\hat{\rho}_0\hat{\sigma}_0^{-1}(s)\right]&=\frac{12}{\nt^2}\\
\label{eq:tr2dK4u0c6Q}-\frac{24}{\nt^2}\trace\left[\hat{\rho}_0\int_0^1s\mathrm{d}s\hat{\sigma}_0^{-1}(s)\hat{\rho}_0\hat{\sigma}_0^{-1}(s)\hat{\rho}_0\hat{\sigma}_0^{-1}(s)\right]&=-\frac{12}{\nt^2}.
\end{align}
Summing
  \eqref{eq:tr2dK4u0c3Q}-\eqref{eq:tr2dK4u0c6Q} yields the second term of 
  \eqref{eq:dK4u0Q}:
\begin{align}
\label{eq:tr2dK4u0evalQ}\trace\left[-6\hat{\rho}_0\int_0^1s\mathrm{d}s\hat{\sigma}_0^{-1}(s)\left.\frac{\mathrm{d}^2\rz}{\mathrm{d} u^2}\right|_{u=0}\hat{\sigma}_0^{-1}(s)\left.\frac{\mathrm{d}^2\rz}{\mathrm{d} u^2}\right|_{u=0}\hat{\sigma}_0^{-1}(s)\right]&=-\frac{12(1+2\nt)}{\nt^2(1+\nt)}+\frac{12}{\nt^2}
\end{align}
Substitution of \eqref{eq:drho4u0} in the third term of \eqref{eq:dK4u0Q} and taking
  the trace yields a sum of five terms, however, the trace is zero for terms comprised of products of states that
  are diagonal in Fock basis and have unequal number of creation and annhihilation operators
  (e.g., 
  $\trace[\hat{\rho}_0\hat{\sigma}_0(s)\hat{a}^2\hat{\rho}_0\hat{\sigma}_0(s)]=0$). The terms with a non-zero trace are as follows:
\begin{align}
\nonumber\trace\left[\hat{\rho}_0\int_0^1\mathrm{d}s\hat{\sigma}_0^{-1}(s)\left.\frac{\mathrm{d}^4\rz}{\mathrm{d} u^4}\right|_{u=0}\hat{\sigma}_0^{-1}(s)\right]&=\frac{12}{\nt^2}\trace\left[\hat{\rho}_0\int_0^1\mathrm{d}s\hat{\sigma}_0^{-1}(s)\hat{\rho}_0\hat{\sigma}_0^{-1}(s)\right]\\
\nonumber&\phantom{=}-\frac{24}{\nt^3}\trace\left[\hat{\rho}_0\int_0^1\mathrm{d}s\hat{\sigma}_0^{-1}(s)\hat{a}\hat{\rho}_0\hat{a}^\dagger\hat{\sigma}_0^{-1}(s)\right]\\
\label{eq:tr3dK4u0Q}&\phantom{=}+\frac{6}{\nt^4}\trace\left[\hat{\rho}_0\int_0^1\mathrm{d}s\hat{\sigma}_0^{-1}(s)\hat{a}^2\hat{\rho}_0(\hat{a}^\dagger)^2\hat{\sigma}_0^{-1}(s)\right].
\end{align}
The first two terms of \eqref{eq:tr3dK4u0Q} can be evaluated using
  \eqref{eq:intidentityQ} and \eqref{eq:intaadaggerQ}:
\begin{align}
\label{eq:tr3dK4u0c1Q}\frac{12}{\nt^2}\trace\left[\hat{\rho}_0\int_0^1\mathrm{d}s\hat{\sigma}_0^{-1}(s)\hat{\rho}_0\hat{\sigma}_0^{-1}(s)\right]&=\frac{12}{\nt^2}\\
\label{eq:tr3dK4u0c2Q}-\frac{24}{\nt^3}\trace\left[\hat{\rho}_0\int_0^1\mathrm{d}s\hat{\sigma}_0^{-1}(s)\hat{a}\hat{\rho}_0\hat{a}^\dagger\hat{\sigma}_0^{-1}(s)\right]&=-\frac{24}{\nt^2}
\end{align}
Since
\begin{align}
\nonumber\int_0^1\mathrm{d}s\hat{\sigma}_0^{-1}(s)\hat{a}^2\hat{\rho}_0(\hat{a}^\dagger)^2\hat{\sigma}_0^{-1}(s)&=\int_0^1\mathrm{d}s\sum_{\sumvar=0}^\infty(\sumvar+1)(n+2)t_{\sumvar+2}(st_\sumvar+(1-s))^{-2}\ket{\sumvar}\bra{\sumvar}\\
\label{eq:intaadagger2Q}&=\frac{\nt^2}{(1+\nt)^2}\sum_{\sumvar=0}^\infty(\sumvar+1)(\sumvar+2)\ket{\sumvar}\bra{\sumvar},
\end{align}
the third term of \eqref{eq:tr3dK4u0Q} is:
\begin{align}
\label{eq:tr3dK4u0c3Q}\frac{6}{\nt^4}\trace\left[\hat{\rho}_0\int_0^1\mathrm{d}s\hat{\sigma}_0^{-1}(s)\hat{a}^2\hat{\rho}_0(\hat{a}^\dagger)^2\hat{\sigma}_0^{-1}(s)\right]&=\frac{12}{\nt^2}.
\end{align}
Summing \eqref{eq:tr3dK4u0c1Q}, \eqref{eq:tr3dK4u0c2Q}, and \eqref{eq:tr3dK4u0c3Q} yields
  the third term of \eqref{eq:dK4u0Q}:
\begin{align}
\label{eq:tr3dK4u0evalQ}\trace\left[\hat{\rho}_0\int_0^1\mathrm{d}s\hat{\sigma}_0^{-1}(s)\left.\frac{\mathrm{d}^4\rz}{\mathrm{d} u^4}\right|_{u=0}\hat{\sigma}_0^{-1}(s)\right]&=0
\end{align}
Summing \eqref{eq:tr1dK4u0evalQ}, \eqref{eq:tr2dK4u0evalQ}, and \eqref{eq:tr3dK4u0evalQ} 
  yields the fourth term in the Taylor series 
  $\frac{1}{4!}\frac{\mathrm{d}^4\kh}{\mathrm{d} u^4}=\frac{1}{2\nt(1+\nt)}$. 
Since $u=\sqrt{1-\eta}b$ and $\nt = \eta\bar{n}_{\rm B}$, we have:
\begin{align}
\label{eq:qreQPSK}D\left(\rz\|\hat{\rho}_{\eta \bar{n}_{\rm B}}\right) = \frac{(1-\eta)^2\bar{n}_{\rm S}^2}{2\eta\bar{n}_{\rm B}(1+\eta\bar{n}_{\rm B})}+o(\bar{n}_{\rm S}^2).
\end{align}
Combining \eqref{eq:qreQPSK} with \eqref{eq:qreineq} (with $\hat{\rho}^W_{1,m}$ set to $\rz$), dropping low order terms, and solving for $\bar{n}_{\rm S}$ yields the proof.
\end{IEEEproof}
\renewcommand{\BPSK}{\rm{b}}
\renewcommand{\rz}{\hat{\rho}_{1,\BPSK}}
\renewcommand{\kh}{\hat{K}_{\BPSK}}
\subsection{Binary phase shift keying}
While BPSK is known to achieve the Holevo capacity
  of (non-covert) communication over lossy thermal noise bosonic channel
  in the low received SNR regime \cite{lacerda17cohstateconstellations},
  here we argue that it is strictly suboptimal for achieving covertness. 
We use the definitions of $u$ and $\nt$ as in Section \ref{sec:QPSK}. We define $\rz = \frac{1}{2}(\ra+\rc)$, where subscript ``b'' stands for BPSK. We evaluate the Taylor series expansion as we did for QPSK. The first and third derivatives of $\rz$ with respect to $u$ evaluated at $u=0$ are zero. The second and fourth derivatives are as follows:
\begin{align}
\label{eq:BPSKdrho2u0}\left.\frac{\mathrm{d}^2\rz}{\mathrm{d}u^2}\right|_{u=0}&=\frac{1}{\nt^2}\left(\hat{a}^2\hat{\rho}_0+2\hat{a}\hat{\rho}_0\hat{a}^\dagger+\hat{\rho}_0(\hat{a}^\dagger)^2\right)-\frac{2}{\nt}\hat{\rho}_0,\\
\nonumber\left.\frac{\mathrm{d}^4\rz}{\mathrm{d}u^4}\right|_{u=0}&=\frac{12\hat{\rho}_0}{\nt^2}-\frac{12}{\nt^3}\left(\hat{a}^2\hat{\rho}_0+2\hat{a}\hat{\rho}_0\hat{a}^\dagger+\hat{\rho}_0(\hat{a}^\dagger)^2\right)\\
\label{eq:BPSKdrho4u0}&\phantom{=}+\frac{1}{\nt^4}\left(\hat{a}^4\hat{\rho}_0+4\hat{a}^3\hat{\rho}_0\hat{a}^\dagger+6\hat{a}^2\hat{\rho}_0(\hat{a}^\dagger)^2+4\hat{a}\hat{\rho}_0(\hat{a}^\dagger)^3+\hat{\rho}_0(\hat{a}^\dagger)^4\right)
\end{align}
Here, $\kh=\rz\log\rz - \rz\log\ro$.
The first three terms of the Taylor series expansion are zero for the BPSK case as their form is similar to the QPSK ones. Let's evaluate the fourth term.
Using Lemma \ref{lemma:dinv}, the fourth derivative of $\kh$ with respect to $u$ evaluated at $u=0$ is:
\begin{align}
\nonumber\left.\frac{\mathrm{d}^4\kh}{\mathrm{d} u^4}\right|_{u=0}&=6\left.\frac{\mathrm{d}^2\rz}{\mathrm{d} u^2}\right|_{u=0}\int_0^1\mathrm{d}s\hat{\sigma}_0^{-1}(s)\left.\frac{\mathrm{d}^2\rz}{\mathrm{d} u^2}\right|_{u=0}\hat{\sigma}_0^{-1}(s)\\
\nonumber&\phantom{=}-6\hat{\rho}_0\int_0^1s\mathrm{d}s\hat{\sigma}_0^{-1}(s)\left.\frac{\mathrm{d}^2\rz}{\mathrm{d} u^2}\right|_{u=0}\hat{\sigma}_0^{-1}(s)\left.\frac{\mathrm{d}^2\rz}{\mathrm{d} u^2}\right|_{u=0}\hat{\sigma}_0^{-1}(s)\\
\label{eq:BPSKdK4u0}&\phantom{=}+\hat{\rho}_0\int_0^1\mathrm{d}s\hat{\sigma}_0^{-1}(s)\left.\frac{\mathrm{d}^4\rz}{\mathrm{d} u^4}\right|_{u=0}\hat{\sigma}_0^{-1}(s).
\end{align}
When evaluating the trace of \eqref{eq:BPSKdK4u0}, we use the fact that $\hat{\sigma}_0(s)$
  is diagonal in Fock basis, and that the trace is zero for terms comprised of states 
  that are diagonal in Fock basis and unequal numbers of creation and annihilation operators,
just as we did in evaluating the trace of the third term of \eqref{eq:tr3dK4u0Q}.
Thus, substitution of \eqref{eq:BPSKdrho2u0} in the first term of \eqref{eq:BPSKdK4u0} and taking
  the trace yields:
\begin{IEEEeqnarray}{Lll}
	\IEEEeqnarraymulticol{3}{l}{
		\nonumber\trace\left[6\left.\frac{\mathrm{d}^2\rz}{\mathrm{d} u^2}\right|_{u=0}\int_0^1\mathrm{d}s\hat{\sigma}_0^{-1}(s)\left.\frac{\mathrm{d}^2\rz}{\mathrm{d} u^2}\right|_{u=0}\hat{\sigma}_0^{-1}(s)\right]
	}\nonumber\\ \qquad
	\nonumber & = & \frac{6}{\nt^4}\trace\left[\hat{a}^2\hat{\rho}_0\int_0^1\mathrm{d}s\hat{\sigma}_0^{-1}(s)\hat{\rho}_0(\hat{a}^\dagger)^2\hat{\sigma}_0^{-1}(s)\right] +\frac{6}{\nt^4}\trace\left[\hat{\rho}_0(\hat{a}^\dagger)^2\int_0^1\mathrm{d}s\hat{\sigma}_0^{-1}(s)\hat{a}^2\hat{\rho}_0\hat{\sigma}_0^{-1}(s)\right]
	\\
	\nonumber & & +\frac{24}{\nt^4}\trace\left[\hat{a}\hat{\rho}_0\hat{a}^\dagger\int_0^1\mathrm{d}s\hat{\sigma}_0^{-1}(s)\hat{a}\hat{\rho}_0\hat{a}^\dagger\hat{\sigma}_0^{-1}(s)\right]-\frac{24}{\nt^3}\trace\left[\hat{a}\hat{\rho}_0\hat{a}^\dagger\int_0^1\mathrm{d}s\hat{\sigma}_0^{-1}(s)\hat{\rho}_0\hat{\sigma}_0^{-1}(s)\right]
	\\
	\label{eq:BPSKtr1dK4u0} & & -\frac{24}{\nt^3}\trace\left[\hat{\rho}_0\int_0^1\mathrm{d}s\hat{\sigma}_0^{-1}(s)\hat{a}\hat{\rho}_0\hat{a}^\dagger\hat{\sigma}_0^{-1}(s)\right] +\frac{24}{\nt^2}\trace\left[\hat{\rho}_0\int_0^1\mathrm{d}s\hat{\sigma}_0^{-1}(s)\hat{\rho}_0\hat{\sigma}_0^{-1}(s)\right].
\end{IEEEeqnarray}
Since $\int_0^1\mathrm{d}s(sq+(1-s))^{-1}(sr+(1-s))^{-1}=\frac{\log\left(\frac{q}{r}\right)}{q-r}$ for $q,r>0$ and $q\neq r$, 
we have:
\begin{align}
\nonumber\int_0^1\mathrm{d}s\hat{\sigma}_0^{-1}(s)\hat{\rho}(\hat{a}^\dagger)^2\hat{\sigma}_0^{-1}(s)&=\int_0^1\mathrm{d}s\sum_{\sumvar=0}^\infty\frac{\sqrt{\sumvar(\sumvar-1)}t_\sumvar\ket{\sumvar}\bra{\sumvar-2}}{(st_\sumvar+(1-s))(st_{\sumvar-2}+(1-s))}\\
\label{eq:BPSKintad2}&=\frac{2\nt^2}{1+2\nt}\log\left(1+\frac{1}{\nt}\right)\sum_{\sumvar=0}^\infty\sqrt{\sumvar(\sumvar-1)}\ket{\sumvar}\bra{\sumvar-2}.
\end{align}
  The second term is obtained similarly to \eqref{eq:BPSKintad2}.
Thus, the first two terms of \eqref{eq:BPSKtr1dK4u0} are:
\begin{align}
\label{eq:BPSKtr1dK4u0c1}\frac{6}{\nt^4}\trace\left[\hat{a}^2\hat{\rho}_0\int_0^1\mathrm{d}s\hat{\sigma}_0^{-1}(s)\hat{\rho}_0(\hat{a}^\dagger)^2\hat{\sigma}_0^{-1}(s)\right]&=\frac{24}{1+2\nt}\log\left(1+\frac{1}{\nt}\right)\\
\label{eq:BPSKtr1dK4u0c2}\frac{6}{\nt^4}\trace\left[\hat{\rho}_0(\hat{a}^\dagger)^2\int_0^1\mathrm{d}s\hat{\sigma}_0^{-1}(s)\hat{a}^2\hat{\rho}_0\hat{\sigma}_0^{-1}(s)\right]&=\frac{24}{1+2\nt}\log\left(1+\frac{1}{\nt}\right)
\end{align}
Comparing \eqref{eq:BPSKtr1dK4u0} and \eqref{eq:tr1dK4u0Q} yields \eqref{eq:BPSKtr1dK4u0c1} and \eqref{eq:BPSKtr1dK4u0c2} as the only terms unique to \eqref{eq:BPSKtr1dK4u0} while the rest are shared. Thus, summing \eqref{eq:BPSKtr1dK4u0c1}, \eqref{eq:BPSKtr1dK4u0c2}, and the shared terms in \eqref{eq:tr1dK4u0evalQ} yields the first term of \eqref{eq:BPSKdK4u0}:
\begin{align}
\label{eq:BPSKtr1dK4u0eval}\trace\left[6\left.\frac{\mathrm{d}^2\rz}{\mathrm{d} u^2}\right|_{u=0}\int_0^1\mathrm{d}s\hat{\sigma}_0^{-1}(s)\left.\frac{\mathrm{d}^2\rz}{\mathrm{d} u^2}\right|_{u=0}\hat{\sigma}_0^{-1}(s)\right]&=\frac{48}{1+2\bar{n}_{\mathrm{T}}}\log\left(1+\frac{1}{\bar{n}_{\mathrm{T}}}\right)+\frac{24(1+2\bar{n}_{\mathrm{T}})}{\bar{n}_{\mathrm{T}}^2(1+\bar{n}_{\mathrm{T}})}-\frac{24}{\bar{n}_{\mathrm{T}}^2}.
\end{align}
Substitution of \eqref{eq:BPSKdrho2u0} in the second term of \eqref{eq:BPSKdK4u0} and taking
  the trace yields:
\begin{IEEEeqnarray}{rCl}
	\IEEEeqnarraymulticol{3}{l}{
		\trace\left[-6\hat{\rho}_0\int_0^1s\mathrm{d}s\hat{\sigma}_0^{-1}(s)\left.\frac{\mathrm{d}^2\rz}{\mathrm{d} u^2}\right|_{u=0}\hat{\sigma}_0^{-1}(s)\left.\frac{\mathrm{d}^2\rz}{\mathrm{d} u^2}\right|_{u=0}\hat{\sigma}_0^{-1}(s)\right]
	}\nonumber\\* \qquad \qquad \qquad
	\nonumber& = & -\frac{6}{\nt^4}\trace\left[\hat{\rho}_0\int_0^1s\mathrm{d}s\hat{\sigma}_0^{-1}(s)\hat{a}^2\hat{\rho}_0\hat{\sigma}_0^{-1}(s)\hat{\rho}_0(\hat{a}^\dagger)^2\hat{\sigma}_0^{-1}(s)\right]
	\\
	\nonumber & & -\frac{6}{\nt^4}\trace\left[\hat{\rho}_0\int_0^1s\mathrm{d}s\hat{\sigma}_0^{-1}(s)\hat{\rho}_0(\hat{a}^\dagger)^2\hat{\sigma}_0^{-1}(s)\hat{a}^2\hat{\rho}_0\hat{\sigma}_0^{-1}(s)\right]
	\\
	\nonumber& & -\frac{24}{\nt^4}\trace\left[\hat{\rho}_0\int_0^1s\mathrm{d}s\hat{\sigma}_0^{-1}(s)\hat{a}\hat{\rho}_0\hat{a}^\dagger\hat{\sigma}_0^{-1}(s)\hat{a}\hat{\rho}_0\hat{a}^\dagger\hat{\sigma}_0^{-1}(s)\right]
	\\
	\nonumber& & +\frac{24}{\nt^3}\trace\left[\hat{\rho}_0\int_0^1s\mathrm{d}s\hat{\sigma}_0^{-1}(s)\hat{\rho}_0\hat{\sigma}_0^{-1}(s)\hat{a}\hat{\rho}_0\hat{a}^\dagger\hat{\sigma}_0^{-1}(s)\right]
	\\
	\nonumber& & +\frac{24}{\nt^3}\trace\left[\hat{\rho}_0\int_0^1s\mathrm{d}s\hat{\sigma}_0^{-1}(s)\hat{a}\hat{\rho}_0\hat{a}^\dagger\hat{\sigma}_0^{-1}(s)\hat{\rho}_0\hat{\sigma}_0^{-1}(s)\right]
	\\
	\label{eq:BPSKtr2dK4u0}& & -\frac{24}{\nt^2}\trace\left[\hat{\rho}_0\int_0^1s\mathrm{d}s\hat{\sigma}_0^{-1}(s)\hat{\rho}_0\hat{\sigma}_0^{-1}(s)\hat{\rho}_0\hat{\sigma}_0^{-1}(s)\right].
\end{IEEEeqnarray}
Since $\int_0^1s\mathrm{d}s(sq+(1-s))^{-2}(sr+(1-s))^{-1}=\frac{r-q+u\log\left(\frac{q}{r}\right)}{q(q-r)^2}$ for $q,r>0$ and $q\neq r$, 
\begin{IEEEeqnarray}{rCl}
	\IEEEeqnarraymulticol{3}{l}{
		\int_0^1s\mathrm{d}s\hat{\sigma}_0^{-1}(s)\hat{a}^2\hat{\rho}_0\hat{\sigma}_0^{-1}(s)\hat{\rho}_0(\hat{a}^\dagger)^2\hat{\sigma}_0^{-1}(s)
	}\nonumber\\* \qquad \qquad
	\nonumber & = & \int_0^1s\mathrm{d}s\sum_{\sumvar=0}^\infty\frac{(\sumvar+1)(\sumvar+2)t_{\sumvar+2}^2\ket{\sumvar}\bra{\sumvar}}{(st_\sumvar+(1-s))^2(st_{\sumvar+2}+(1-s))}
	\\
	\label{eq:BPSKtr2dK4u0c1calc} & = & \left(\frac{2\nt^4}{(1+2\nt)^2}\log\left(1+\frac{1}{\nt}\right)-\frac{\nt^4}{(1+\nt)^2(1+2\nt)}\right)\sum_{\sumvar=0}^\infty(\sumvar+1)(\sumvar+2)\ket{\sumvar}\bra{\sumvar}.
\end{IEEEeqnarray}
  The second term is evaluated similarly to \eqref{eq:BPSKtr2dK4u0c1calc}.
Thus, the first two terms of \eqref{eq:BPSKtr2dK4u0} are:
\begin{align}
\label{eq:BPSKtr2dK4u0c1}-\frac{6}{\nt^4}\trace\left[\hat{\rho}_0\int_0^1s\mathrm{d}s\hat{\sigma}_0^{-1}(s)\hat{a}^2\hat{\rho}_0\hat{\sigma}_0^{-1}(s)\hat{\rho}_0(\hat{a}^\dagger)^2\hat{\sigma}_0^{-1}(s)\right]&=-\frac{24(1+\nt)^2}{(1+2\nt)^2}\log\left(1+\frac{1}{\nt}\right)+\frac{12}{1+2\nt}\\
\label{eq:BPSKtr2dK4u0c2}-\frac{6}{\nt^4}\trace\left[\hat{\rho}_0\int_0^1s\mathrm{d}s\hat{\sigma}_0^{-1}(s)\hat{\rho}_0(\hat{a}^\dagger)^2\hat{\sigma}_0^{-1}(s)\hat{a}^2\hat{\rho}_0\hat{\sigma}_0^{-1}(s)\right]&=\frac{24\nt^2}{(1+2\nt)^2}\log\left(1+\frac{1}{\nt}\right)-\frac{12}{1+2\nt}.
\end{align}
Comparing \eqref{eq:BPSKtr2dK4u0} and \eqref{eq:tr2dK4u0Q} yields \eqref{eq:BPSKtr2dK4u0c1} and \eqref{eq:BPSKtr2dK4u0c2} as the only terms unique to \eqref{eq:BPSKtr2dK4u0} while the rest are shared. Summing \eqref{eq:BPSKtr2dK4u0c1}, \eqref{eq:BPSKtr2dK4u0c2}, and the shared terms in \eqref{eq:tr2dK4u0evalQ} yields the second term of \eqref{eq:BPSKdK4u0}:
\begin{IEEEeqnarray}{rCl}
	\IEEEeqnarraymulticol{3}{l}{
		\nonumber\trace\left[-6\hat{\rho}_0\int_0^1s\mathrm{d}s\hat{\sigma}_0^{-1}(s)\left.\frac{\mathrm{d}^2\rz}{\mathrm{d} u^2}\right|_{u=0}\hat{\sigma}_0^{-1}(s)\left.\frac{\mathrm{d}^2\rz}{\mathrm{d} u^2}\right|_{u=0}\hat{\sigma}_0^{-1}(s)\right]
	}\nonumber\\* \qquad \qquad \qquad \qquad
	\label{eq:BPSKtr2dK4u0eval} & = & -\frac{24}{1+2\nt}\log\left(1+\frac{1}{\nt}\right)-\frac{12(1+2\nt)}{\nt^2(1+\nt)}+\frac{12}{\nt^2}.
\end{IEEEeqnarray}
Substitution of \eqref{eq:BPSKdrho4u0} in the third term of \eqref{eq:BPSKdK4u0} and taking
  the trace yields:
\begin{align}
\nonumber\trace\left[\hat{\rho}_0\int_0^1\mathrm{d}s\hat{\sigma}_0^{-1}(s)\left.\frac{\mathrm{d}^4\rz}{\mathrm{d} u^4}\right|_{u=0}\hat{\sigma}_0^{-1}(s)\right]&=\frac{12}{\nt^2}\trace\left[\hat{\rho}_0\int_0^1\mathrm{d}s\hat{\sigma}_0^{-1}(s)\hat{\rho}_0\hat{\sigma}_0^{-1}(s)\right]\\
\nonumber&\phantom{=}-\frac{24}{\nt^3}\trace\left[\hat{\rho}_0\int_0^1\mathrm{d}s\hat{\sigma}_0^{-1}(s)\hat{a}\hat{\rho}_0\hat{a}^\dagger\hat{\sigma}_0^{-1}(s)\right]\\
\label{eq:BPSKtr3dK4u0}&\phantom{=}+\frac{6}{\nt^4}\trace\left[\hat{\rho}_0\int_0^1\mathrm{d}s\hat{\sigma}_0^{-1}(s)\hat{a}^2\hat{\rho}_0(\hat{a}^\dagger)^2\hat{\sigma}_0^{-1}(s)\right].
\end{align}
Comparison of \eqref{eq:BPSKtr3dK4u0} and \eqref{eq:tr3dK4u0Q} shows that they are equal. Since \eqref{eq:tr3dK4u0evalQ} shows this term to be zero, the third term of \eqref{eq:BPSKdK4u0} is zero. 
Summing \eqref{eq:BPSKtr1dK4u0eval} and \eqref{eq:BPSKtr2dK4u0eval} yields the fourth term in the Taylor series:
\begin{align}
\label{eq:BPSKtseval}\frac{1}{4!}\frac{\mathrm{d}^4\kh(u)}{\mathrm{d} u^4}&=\frac{1}{2\nt(1+\nt)}+\frac{1}{1+2\nt}\log\left(1+\frac{1}{\nt}\right)
\end{align}
The QRE for BPSK is as follows:
\begin{align}
\label{eq:qrebpsk}D\left(\hat{\rho}_{1,b}\|\hat{\rho}_{\eta \bar{n}_{\rm B}}\right) = (1-\eta)^2\bar{n}_{\rm S}^2\left[\frac{1}{2\eta\bar{n}_{\rm B}(1+\eta\bar{n}_{\rm B})} + \frac{1}{1+2\eta\bar{n}_{\rm B}}\log\left(1+\frac{1}{\eta\bar{n}_{\rm B}}\right)\right]+o(\bar{n}_{\rm S}^2).
\end{align}
This is strictly larger than $D\left(\hat{\rho}_{1,q}\|\hat{\rho}_{\eta \bar{n}_{\rm B}}\right)$ in \eqref{eq:qreQPSK}. Therefore, to maintain Criterion \ref{crit:covcritRE}, $\bar{n}_{\rm S}$ must be set strictly less than the optimal value in \eqref{eq:c_det}. 
\subsection{Use of practical transmitters and arbitrary codes}\label{sec:practical}
Typical optical transmitters operate at a constant mean photon number per mode
  $\bar{n}_{\rm S}$, and much of coding theory assumes that $\bar{n}_{\rm S}$ is 
  independent from $n$.
However, covertness requires $\bar{n}_{\rm S}$ to decay with $n$.
We address this by modifying the construction of the secret random sequence 
  described in Section \ref{sec:discrete_coding}.
First, Alice and Bob secretly select a subset of modes $\mathcal{S}$ for
  communication by flipping a random coin $n$ times with probability of heads 
  $\tau$.
The $k^{\rm th}$ mode is chosen if the $k^{\rm th}$ flip is heads. 
They then generate the secret random sequence as described in Section
  \ref{sec:discrete_coding}, and use a public code on the modes
  in set $\mathcal{S}$ of expected size $E[|\mathcal{S}|]=\tau n$.
Let Alice use the coherent state QPSK modulation.
Since Willie does not have $\mathcal{S}$, when she transmits, he observes 
  $\hat{\rho}_{1,\tau}=(1-\tau)\hat{\rho}_0+\tau\hat{\rho}_{1,\mathrm{q}}$ on
  each of $n$ modes, with $\hat{\rho}_0$ and 
  $\hat{\rho}_{1,\mathrm{q}}$ defined in Section \ref{sec:optimalQPSK}.
Note that $\frac{\mathrm{d}^n\hat{\rho}_{1,\tau}}{\mathrm{d}u^n}=\tau\frac{\mathrm{d}^n\hat{\rho}_{1,\mathrm{q}}}{\mathrm{d}u^n}$.
Replacing 
  $\left.\frac{\mathrm{d}^n\hat{\rho}_{1,\mathrm{q}}}{\mathrm{d}u^n}\right|_{u=0}$
  with
  $\left.\frac{\mathrm{d}^n\hat{\rho}_{1,\tau}}{\mathrm{d}u^n}\right|_{u=0}$ in
  \eqref{eq:dK4Q} yields:
\begin{align}
\label{eq:qreQPSKtau}D\left(\hat{\rho}_{1,\tau}\|\hat{\rho}_{\eta \bar{n}_{\rm B}}\right) = \frac{(1-\eta)^2\tau^2\bar{n}_{\rm S}^2}{2\eta\bar{n}_{\rm B}(1+\eta\bar{n}_{\rm B})}+o(\tau^2\bar{n}_{\rm S}^2).
\end{align}
We discard low order terms, fix $\bar{n}_{\rm S}$, and solve for $\tau$
  that maintains Criterion \ref{crit:covcritRE}:
\begin{align}
\label{eq:tau}
\tau&=\frac{\sqrt{2\eta\bar{n}_{\rm B}(1+\eta\bar{n}_{\rm B})}}{(1-\eta)\bar{n}_{\rm S}}\sqrt{\frac{\delta_{\rm QRE}}{n}}.
\end{align}
This method was used in a covert communication experiment described in
  \cite{bash15covertbosoniccomm}.
When a Holevo-achieving code is used (with a constant $\bar{n}_{\rm S}$) it enables
  the achievability of the ultimate limit of covert communication over the 
  bosonic channel in expectation, as described in Section \ref{sec:coding}.
We also note that it requires $\mathcal{O}(\sqrt{n}\log n)$ bits of shared
  secret \cite[App.]{bash13squarerootjsacnonote}.
We conjecture, based on the results for classical channels \cite{bloch15covert},
  that at most $\mathcal{O}(\sqrt{n})$ shared secret bits are needed for
  reliable covert communication under any conditions on Alice's channels
  to Bob and Willie.
However, the perspective methods to achieve this scaling (e.g., extension of
  \cite{zhang16covertcodes} to arbitrary channel conditions) are impractically
  complex.
We offer simplicity and robustness of existing codes at a mere 
  $\log n$ factor increase in shared secret size, which is an acceptable trade-off
  in many  applications given significantly lower power
  consumption of flash memory vs.~computers.
\section{Coding for covert communication over bosonic channel}
\label{sec:coding}
Constant $c_{\rm rel}$ determines the number of covert bits that are
  reliably transmissible over the bosonic channel.
Here we provide the lower and upper bounds, show how the latter can be met in
  expectation, and offer a roadmap to the complete characterization of 
  $c_{\rm rel}$ in the future work.
The lower bound $c_{\rm rel}$ is straightforward: restrict Bob to 
  a heterodyne receiver, yielding a classical AWGN channel that is characterized in 
  \cite[Eq.~1.3]{guha08phdthesis}.
We then employ the known results \cite{bloch15covert,wang15covert} 
  to obtain 
  $c_{\rm rel}\geq \eta((1-\eta)\bar{n}_{\rm B})^{-1}$.
For the upper bound, observe that the Holevo capacity of the bosonic channel is
  additive.
Thus, the number of covert bits 
  that can be transmitted reliably over such channel with transmissivity $\eta$ and 
  mean thermal noise photon number $\bar{n}_{\rm B}$ is 
  $M=nB(\bar{n}_{\rm S};\eta,\bar{n}_{\rm B})$, where 
  $B(\bar{n}_{\rm S};\eta,\bar{n}_{\rm B})$ is 
  the number of transmissible bits using $\bar{n}_{\rm S}$ photons per mode.
The Holevo capacity of the lossy thermal noise bosonic channel upper bounds 
  $B(\bar{n}_{\rm S};\eta,\bar{n}_{\rm B})\leq\chi(\bar{n}_{\rm S};\eta,\bar{n}_{\rm B})$, 
  and has been characterized for non-covert scenarios \cite{Sha05}.
Since $\bar{n}_{\rm S}$ is small for large $n$, we can upper-bound
  $c_{\rm rel}\leq c_{\rm rel,\chi}$
  by the first Taylor series term of $\chi(\bar{n}_{\rm S};\eta,\bar{n}_{\rm B})$ 
  \cite[Eq.~(10)]{Sha05} expanded at $\bar{n}_{\rm S}=0$: 
  $c_{\rm rel,\chi}= \eta\log\left(1+((1-\eta)\bar{n}_{\rm B})^{-1}\right)$.
This bound can be achieved in expectation using the coin flip method
  described in Section \ref{sec:practical} by setting $\bar{n}_{\rm S}$ to
  a constant and employing a Holevo capacity achieving code.
Holevo--Schumacher--Westmoreland (HSW) theorem 
  \cite[Sec.~20.3.1]{wilde16quantumit2ed} allows the construction of such code 
  over the subset of modes chosen by the coin flip
  process since $\bar{n}_{\rm S}$ is constant.
A polar code \cite{wilde13polarcodes,nasser18polar} over QPSK
  constellation achieves the Holevo capacity 
  at low signal to noise ratio (SNR) \cite{lacerda17cohstateconstellations}.
Thus $\mathrm{E}[M]=\sqrt{n}\delta c_{\rm cov}c_{\mathrm{rel},\chi}$, with the
  expectation taken over the binomial random variable $\mathcal{B}(\tau,n)$,
  where $\tau$ is defined in \eqref{eq:tau}.
However, we conjecture that the $c_{\rm rel,\chi}$ is achievable in general.
In covert communication 
  $\bar{n}_{\rm S}=\delta c_{\rm cov}/\sqrt{n}$, and this dependence of 
  $\bar{n}_{\rm S}$ on $n$ complicates the application of HSW theorem.
Classical results \cite{bloch15covert,wang15covert} overcome this problem using
  information spectrum methods and resolvability.
The quantum predecessors of these classical methods have been used
  to strengthen the capacity results for classical-quantum channels 
  \cite{hayashi2003general,nagaoka2007information}.
Unfortunately, their use in bosonic channel setting has been limited because
  of their dependence on the finite dimensionality of the Hilbert space for 
  the output quantum states, while the output of the lossy thermal noise bosonic
  channel lives in an infinitely-dimensional Hilbert space.
That being said, one could conceivably adapt the proofs in
  \cite{hayashi2003general,nagaoka2007information}
  to the special case of finite output state constellation, which is indeed
  what we showed to be optimal under the covertness constraint.
\section{Conclusion}
\label{sec:discussion}
Our main objective was to establish the theoretical groundwork 
  necessary for implementation of quantum-secure covert communication over  
  practical channels.
Hence we focused on the bosonic channel model, which is the underlying 
  quantum-mechanical description of many significant communication channels
  (including optical, microwave, and RF).
We have characterized the constant $c_{\rm cov}$ in the expression for mean
  photon number per mode $\bar{n}_{\rm S}=\delta c_{\rm cov}/\sqrt{n}$ in the SRL
  for the lossy thermal noise bosonic channel by proving the converse that matched
  a previous achievability result \cite[Th.~2]{bash15covertbosoniccomm}.
We proved that coherent state QPSK modulation carries the maximum mean photon
  number that covertness requirement allows, and showed that it
  yields optimal covert throughput over the bosonic channel in expectation,
  provided that QPSK modulation achieves Holevo capacity (which it does 
  at low SNR \cite{lacerda17cohstateconstellations}).
While we left the full characterization of covert channel code for future work, 
  we believe that our result opens a clear path to use
  polar codes for quantum-secure covert communications, as the explicit successive 
  cancellation decoder structure is known for discrete constellations
  \cite{wilde13polarcodes,nasser18polar}.
More importantly, we showed that we can ensure quantum-secure covertness using
  practical systems that employ constant-amplitude lasers and coherent receivers.
There are many avenues for future research.
Here we assume that the adversary knows when the transmission may
  start and end, as well as its center frequency and bandwidth.
Asynchronous covert communication lifts these assumptions.
It has been shown that the number of reliable covert bits increases substantially
  in classical AWGN scenario \cite{bash14timingisit,bash16timingtwcnonote}.
This result was later extended to discrete memoryless channels (DMCs)
  \cite{arumugam16async}.
Bosonic channel is a natural setting for further exploration of this topic.
While QRE is mathematically convenient, the trace distance carries more
  operational significance from its direct relationship to the minimum detection
  error probability.
Extension of \cite{tahmasbi19covertdmc2ndorder} to quantum 
  systems would enable analysis of covert communication that is quantum-secure 
  under Criterion \ref{crit:covcritP}.
It might also reveal a path to the evaluation of second-order constants for 
  covert communications over the bosonic channel.
Also, the characterization of covert communication over arbitrary quantum 
  channels has been elusive.
While the achievability was proven in \cite{azadeh16quantumcovert-isitarxiv}
  by extending the techniques of \cite{bloch15covert} to finite-dimensional
  memoryless quantum channels (modeled by trace-preserving completely positive
  maps), the known converse is restricted to product state transmission.
Recent result \cite{tahmasbi19covertqkd} on covert QKD opens 
  a new perspective on this problem.
Finally, optical receiver designs for quantum-optimal state discrimination are not 
  known beyond binary pure state discrimination \cite{dolinar73}.
For discriminating a constellation of size $m>2$, the same physical resources
  that achieve optimal $m=2$ state discrimination (linear optics, laser local 
  oscillator, photon detector, and electro-optical feedback) do not suffice
  \cite{dasilva13discrim}.
For mixed states such as displaced thermal states, the optimal receiver design is 
  not known even for the binary case.
We expect a similar quantum resource divide in this case as in the pure
  state case, and the separation in discriminability between BPSK
  and QPSK that we showed may lead to new insights into this problem.

\end{document}